\documentclass{aa}  
\usepackage{graphicx}
\usepackage{natbib}
\usepackage[colorlinks=true, allcolors=blue]{hyperref}
\usepackage{lscape}

\usepackage{txfonts}

\begin{document}

   \title{Expectations for time-delay measurements in active galactic nuclei with the Vera Rubin Observatory\thanks{on behalf of the LSST-AGN Science Collaboration}}


    \author{Bozena Czerny \inst{1}
          \and
          Swayamtrupta Panda \inst{2}$^{\thanks{CNPq Fellow}}$
          \and
          Raj Prince \inst{1}
          \and
          Vikram Kumar Jaiswal \inst{1}
          \and 
          Michal Zaja\v{c}ek \inst{3}
          \and
          Mary Loli Martinez Aldama \inst{4}
          \and 
          Szymon Koz\l{}owski \inst{5}
          \and
          Andjelka B. Kovacevic \inst{6}
          \and
          Dragana Ilic \inst{6,7}
          \and 
          Luka \v C. Popovi\'c \inst{6,8}
          \and
          Francisco Pozo Nu\~nez \inst{9}
          \and
          Sebastian F. H\" onig \inst{10}
          \and 
          William N. Brandt \inst{11}
          }
  \institute{Center for Theoretical Physics, Polish Academy of Sciences, Al. Lotnik\'ow 32/46, 02-668 Warsaw, Poland
   \and Laborat\'orio Nacional de Astrof\'isica - MCTI, R. dos Estados Unidos, 154 - Na\c{c}\~oes, Itajub\'a - MG, 37504-364, Brazil
   \and Department of Theoretical Physics and Astrophysics, Faculty of Science, Masaryk University, Kotlarska 2, CZ-61137, Brno, Czech Republic
   \and Departamento de Astronomia, Universidad de Chile, Camino del Observatorio 1515, Santiago, Casilla 36-D, Chile
   \and Astronomical Observatory, University of Warsaw, Al.Ujazdowskie 4, PL-00-478 Warszawa, Poland
   \and University of Belgrade-Faculty of Mathematics, Department of astronomy, Studentski trg 16 Belgrade, Serbia
   \and {Humboldt Research Fellow, Hamburger Sternwarte, Universit{\"a}t Hamburg, Gojenbergsweg 112, 21029 Hamburg, Germany}
 \and Astronomical Observatory, Volgina 7, 11160 Belgrade, Serbia
   \and Astroinformatics, Heidelberg Institute for Theoretical Studies, Schloss-Wolfsbrunnenweg 35, 69118 Heidelberg, Germany
   \and School of Physics \& Astronomy, University of Southampton, Southampton SO17 1BJ, UK
   \and Department of Astronomy and Astrophysics, 525 Davey Lab, The Pennsylvania State University, University Park, PA 16802, USA
             }

   \date{Received ???; accepted ???}

 
  \abstract
   {The Vera Rubin Observatory will provide an unprecedented set of time-dependent observations of the sky. The planned  Legacy Survey of Space and Time (LSST), operating  for ten years,  will provide dense light curves for thousands of active galactic nuclei (AGN) in deep drilling fields (DDFs) and less dense light curves for millions of AGN from the main survey (MS).}
   {We model the prospects for measuring the time delays for the AGN emission lines with respect to the continuum, using these data.}
   {We modeled the artificial light curves using the Timmer-K\" onig algorithm. We used the exemplary cadence to sample them (one for the MS and one for the DDF), we supplement light curves with the expected contamination by the strong emission lines (H$\beta$, Mg II, and CIV, as well as with Fe II pseudo-continuum and the starlight). We chose suitable photometric bands that are appropriate for the redshift and compared the assumed line time-delay with the recovered time delay for 100 statistical realizations of the light curves.}
   {We show that time delays for emission lines can be well measured from the main survey for the bright tail of the quasar distribution (about 15\% of all sources) with an accuracy within 1$\sigma$ error. For the DDF, the results for fainter quasars are also reliable when the entire ten years of data are used. There are also some prospects to measure the time delays for the faintest quasars at the lowest redshifts from the first two years of data, and possibly even from the first season. The entire quasar population will allow us to obtain results of apparently high accuracy, but in our simulations, we see a systematic offset between the assumed and recovered time delay that depends on the redshift and source luminosity. This offset will not disappear even in the case of large statistics. This problem might affect the slope of the radius-luminosity relation and cosmological applications of quasars if no simulations are performed that correct for these effects.
   }
   {}


   \maketitle

\section{Introduction}

The Vera C. Rubin Observatory and its Legacy Survey of Space and Time \citep[LSST; ][]{LSST_Ivezic_etal_2019} will provide an unprecedented amount of data in many fields and will thus revolutionize our view of the Universe. Observations will start relatively soon, most likely in mid-2024. Optimizing the output from these data, particularly from the first stages of its operation, is therefore extremely important. Among numerous results, LSST will provide up to ten million quasars. This will open a path to massive reverberation monitoring of active galactic nuclei (AGN) in a range of redshifts  from 0 up to 7 \citep{shen_X_2020, Kovacevic_etal_2021}.

The current description of the Vera Rubin Observatory can be found in \citet{LSST_Ivezic_etal_2019}. General expectations of the LSST discoveries in the field of AGN were discussed by \citet{brandt2018}, and specific predictions for the mapping of AGN accretion disks with the LSST were presented by \citet{LSST_Kovacevic2022} and \citet{pozo2022}. The prospects for the continuum time-delays from accretion disks were also discussed by \citet{Yu_etal_2020}. In the current paper, we aim to assess the accuracy with which emission-line time delays can be measured using the four photometric bands of the LSST. Broad emission lines are characteristic for almost all bright AGN \citep[see, e.g.,][for a suitable review]{Krolikbook}. The exact location and geometry of the corresponding region (known as the broadline region, BLR) is generally unresolved, except for the most recent observations of three AGN in the infrared domain with the use of the GRAVITY Very Large Telescope Interferometer (VLTI) (3C 273; \citealt{GRAVITY_3C273}, NGC 3783; \citealt{GRAVITYNGC3783}, and IRAS 09149-6206; \citealt{GRAVITY_iras}). For the remaining objects, we rely on reverberation monitoring in order to understand the structure and dynamics of their BLRs.

Reverberation mapping is a well-established observational technique that has been extensively used to study the inner structure of AGN on subparsec length scales, which are typically beyond the resolution limits of current telescopes. It effectively replaces the spatial resolution with a time resolution that needs to be adjusted according to the spatial scale of interest \citep[see, e.g.,][for a review and references therein]{2021iSci...24j2557C}. The idea was proposed in 1982 by \citet{1982ApJ...255..419B}, who showed that the line emission produced in the BLR follows the same variability pattern as the continuum emission from the disk, but is delayed by the light travel time between the disk and the BLR. The first systematic observational results were published in 1990 to 1993, presenting the results from the ground-based campaigns as well as from the International Ultraviolet Explorer (IUE) for the bright nearby AGN \citep{maoz1990,netzer1990,peterson_ark_120_1991,peterson_5548_1991,1993PASP..105..247P}. One of the main results of reverberation is the measured time lag, which can be used to estimate the black hole mass. This technique has successfully measured the BH mass of many low-redshift ($z < 0.5$)  AGN through intense monitoring \citep{kaspi2000,peterson2004, 2009ApJ...705..199B, 2015PASP..127...67B,dupu2015,dupu2016,dupu2018}. The results including higher redshifts came later, based on the Mg II and CIV lines, because they required longer monitoring \citep{shen2016,grier2017,lira2018,czerny2019,zajacek2020,kaspi2021,zajacek2021}.
The radius of the BLR measured from the time delays and the monochromatic absolute luminosity of reverberated quasars, and in general in all AGN, are significantly correlated, which is referred to as the radius-luminosity (RL) relation \citep{bentz2013}. Recently, the standard RL relation has been proposed for cosmological applications by \citet{watson2011}, \citet{Haas2011}, and \citet{czerny2013}. The development has been encouraging, although the cosmological constraints are not yet tight because only a few AGN were studied, and the scatter in individual  measurements is also large \citep{MaryLoli2019, Panda_etal_2019, Acta2021, zajacek2021, khadka2021, khadka2022, 2023MNRAS.522.1247K, cao2022, Panda_Marziani_2022}.  

The LSST will provide photometric light curves for many AGN, but the conversion from photometry to line time-delays is not simple and accurate, and it is important to estimate the reliability of the measurements depending on the source luminosity, redshift, and available cadence. We thus created a simulation tool that allowed us to optimize the success of the measurements for the whole survey, the first two years, and the first year, and we considered both the main survey and the planned deep drilling fields.

In  this  work  we  present the prospects for the line delay measurements with LSST data. In Section~\ref{sect:method} we describe the numerical code we created for that purpose. In Section~\ref{sect:results} we show the code predictions based on the planned observational cadence. Methodology limitations are addressed in Section~\ref{sect:discussion}, and a number of additional issues are discussed in the Appendix. Main results are summarized in Section~\ref{sect:conclusions}.

\section{Model}
\label{sect:method}


We tested the prospects for the time-delay determination of broad emission lines with respect to the continuum using the synthetic light curves that were probed according to the  operation simulator (OpSim) provided by the VRO-LSST data management team\footnote{The results of the OpSim runs are hosted on the publicly available website: \href{http://astro-lsst-01.astro.washington.edu:8080}{http://astro-lsst-01.astro.washington.edu:8080}}. 


We modeled the continuum variability and the contribution of the emission lines to the photometric bands. This allowed us to simulate the accuracy of the recovery of the broad emission line delays. In this way, we also optimized the future modeling effort by the choice of prospective sources at each stage for the duration of the project. In our simulations, we assumed that the redshift of the studied source, $z$, can be estimated. This allowed us in principle to derive the absolute value of the monochromatic flux in one of the photometric bands from the measured magnitude. In the current version of the program, we simply assumed the value of the absolute monochromatic flux as one of the parameters because we aim to test the possibility of the time-delay determination, and not at a determination of the cosmological parameters from the measured time delay. If only a photo-z is available, the delay measurement is still possible, but it introduces additional errors \citep{Science_Book_2009, LSST_Ivezic_etal_2019}. We discuss this issue in Sect. 4. 

The presented modeling is based on stochastic light curves, which offer a good representation of the AGN variability, and we present the method used in our simulations in Section~\ref{sect:curve_reference}. We stress that 
because the created light curves have a random character, we always calculated 100 realizations of the full process described in the next subsections for a single set of parameters to assess the accuracy of the time-delay modeling. A preliminary version of the results from our code can be found in \citet{Panda_etal_2019}. 

   \begin{table*}  
   \centering
      \caption{\label{tab:parameters} Model parameters of pairs of photometric light curves.}
         \begin{tabular}{lcr}
            \hline
            \noalign{\smallskip}
           Parameter & notation & default \\
                     &          &    values\\
            \noalign{\smallskip}
            \hline
            \hline
            redshift & z &  \\
            photometric error & P &  0.001 mag \\
            spectral slope $F_{\nu} \propto \nu^{\alpha}$ & $\alpha$ &  0\\
            equivalent width of H$\beta$ & EW(H$\beta$) & 150 \AA\\
            equivalent width of Mg II & EW(MgII) & 47 \AA\\
            equivalent width of CIV & EW(CIV)& 45 \AA \\
            equivalent width of Fe II  & EW(FeII) & 50 \AA \\
            line width & FWHM & 4000 km s$^{-1}$ \\
             dispersion velocity FeII & $\sigma_{FeII}$ &900 km s$^{-1}$ \\
             starlight normalization at 280 nm & Astar & 0.258 \\
             number of statistical realizations & N &  100 \\
            low frequency power spectrum slope & $\alpha_1$ & 0 \\
            low frequency break & $fb_1$ & $3.7 \times 10^{-10}$ Hz\\
            mid-frequency slope & $\alpha_2$ & 1.2 \\
            high frequency break & $fb_2$ & $5.8 \times 10^{-9}$ Hz\\
            high frequency slope & $\alpha_3$ & 2.5 \\
            sampling rate & $\Delta T$ &  1 day \\ 
            total duration of the TK light curve & T & $10^9$ s\\
            assumed variability level & $F_{var}$ & 0.3  \\
            offset in the R-L relation & $\beta$ & 1.573 \\
            slope in the R-L relation & $\gamma$ & 0.5 \\
            width of the BLR Gaussian response &$\sigma_{BLR}$ & 0.1 $\tau$\\ 
            curve subtraction coefficient & $\epsilon_{min}$ & 0.85\\
            curve subtraction coefficient & $\epsilon_{max}$ & 1.15 \\
            no of subtraction sampling  & $n_{\epsilon}$ & 10 \\
            \hline
        \end{tabular}
\end{table*}

\subsection{Choice of suitable photometric bands}

We considered the $g$, $r$, $i$, and $z$ bands as particularly suitable for line-delay measurements because their efficiency is high. We selected only three strong emission lines for the tentative time-delay measurements: H$\beta$, Mg II, and CIV, because they have a record of showing radius-luminosity relation from time-delay measurements (e.g., \citealt[][]{bentz2013} for H$\beta$; e.g., \citealt[][]{homayouni2020,zajacek2021} for Mg II; and, e.g., \citealt[][]{cao2022} for CIV). Since the line should be well within the border of the band edges, we limited the position of the line to 410 - 530 nm (in $g$ band), 570 - 670 nm (in $r$ band), 710-800 nm (in $i$ band), and 830-910 nm (in $z$ band). Modeling photometric reverberation mapping data shows that if the BLR is dominated by circular Keplerian orbits, a symmetric cut of the line wings can lead to an overestimation of the delay by less than 5\%. In the case of asymmetric half-line coverage, the bias of the delay is less than 1\% (see \citealt{pozo2014}). For the line center in the rest frame, we assumed 486.2721 nm (H$\beta$) and 154.90 nm (CIV), and the Mg II line was modeled as a doublet with the mean position 0.5(279.635 + 280.353) nm. If any of the redshifted lines fit any of the favored spectral regions, this line and this photometric band were selected for further modeling as a line-contaminated band. The near band was selected as an uncontaminated band for further processing, and all this was performed automatically. When none of the lines fit into one of the bands, then no time delay recovery was performed for this redshift. 

   \begin{table}  
   \centering
      \caption{\label{tab:bands} {Choice of bands in the standard automatic setup.}}
         \begin{tabular}{lrrr}
            \hline
            \noalign{\smallskip}
           Redshift range & emission & line band & continuum band \\
                     &   line       &    \\
            \noalign{\smallskip}
            \hline
            \hline
      0 - 0.09 & H$\beta$ & g & r\\
      0.173 - 0.377 & H$\beta$ & r & i\\
      0.464 - 0.892 & Mg II & g & r\\
      1.036  - 1.392 & Mg II & r & i\\
      1.647 - 2.421 & CIV & g & r\\
      2.680 - 3.325 & CIV & r & i\\      
            \hline
        \end{tabular}
\end{table}

The choice of a suitable band for the line emission depends on the redshift. For some redshifts, two bands could be considered (e.g., for redshifts $z \sim 0.5,$ both H$\beta$ and Mg II can be studied). Moreover, the choice of the second uncontaminated band is usually not unique as it could be a shorter or longer band next to the one with the line. We usually studied sets of redshifts in an automatic code, in which case, we predefined these setups for all the simulations. We  list our choice in Table~\ref{tab:bands}, but this can be modified in the code when needed. Using the z-band, we could extend the studied range to higher redshifts, but higher-redshift quasars are more likely to be affected by broad absorption lines and at still higher redshifts, by the Ly$\alpha$ forest. We therefore extended our standard plots only up to redshift 3.5. Table~\ref{tab:bands} shows that for some redshifts, we do not have a suitable choice of lines to follow because the potential line is too close to the edge of the photometric band \citep[see][for more details]{Panda_etal_2019}.

\subsection{Artificial spectrum and line contamination of the photometric bands}
\label{sect:contamination}

\begin{figure}  
\centering
\includegraphics[width=\columnwidth]{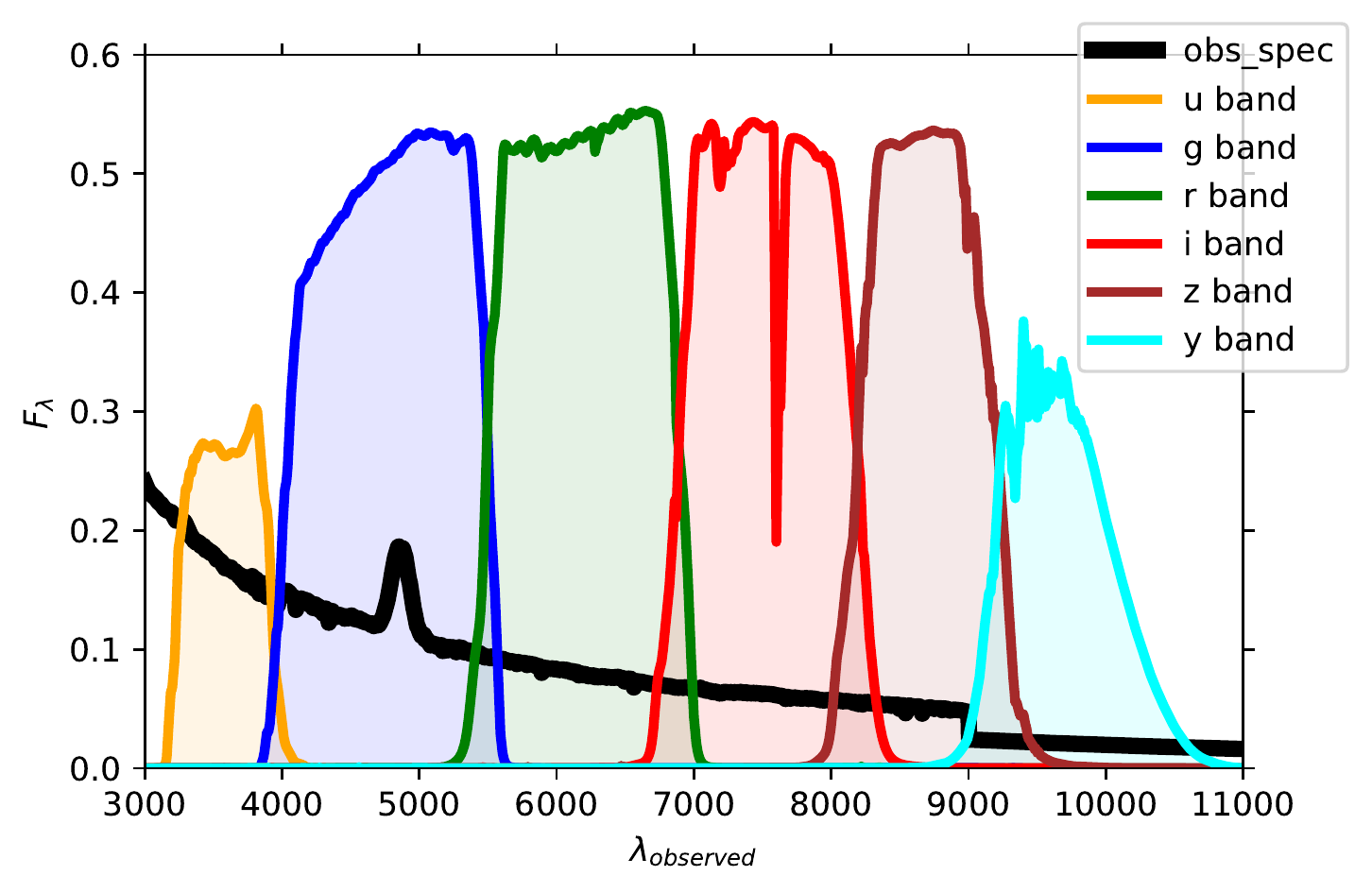}
\includegraphics[width=\columnwidth]{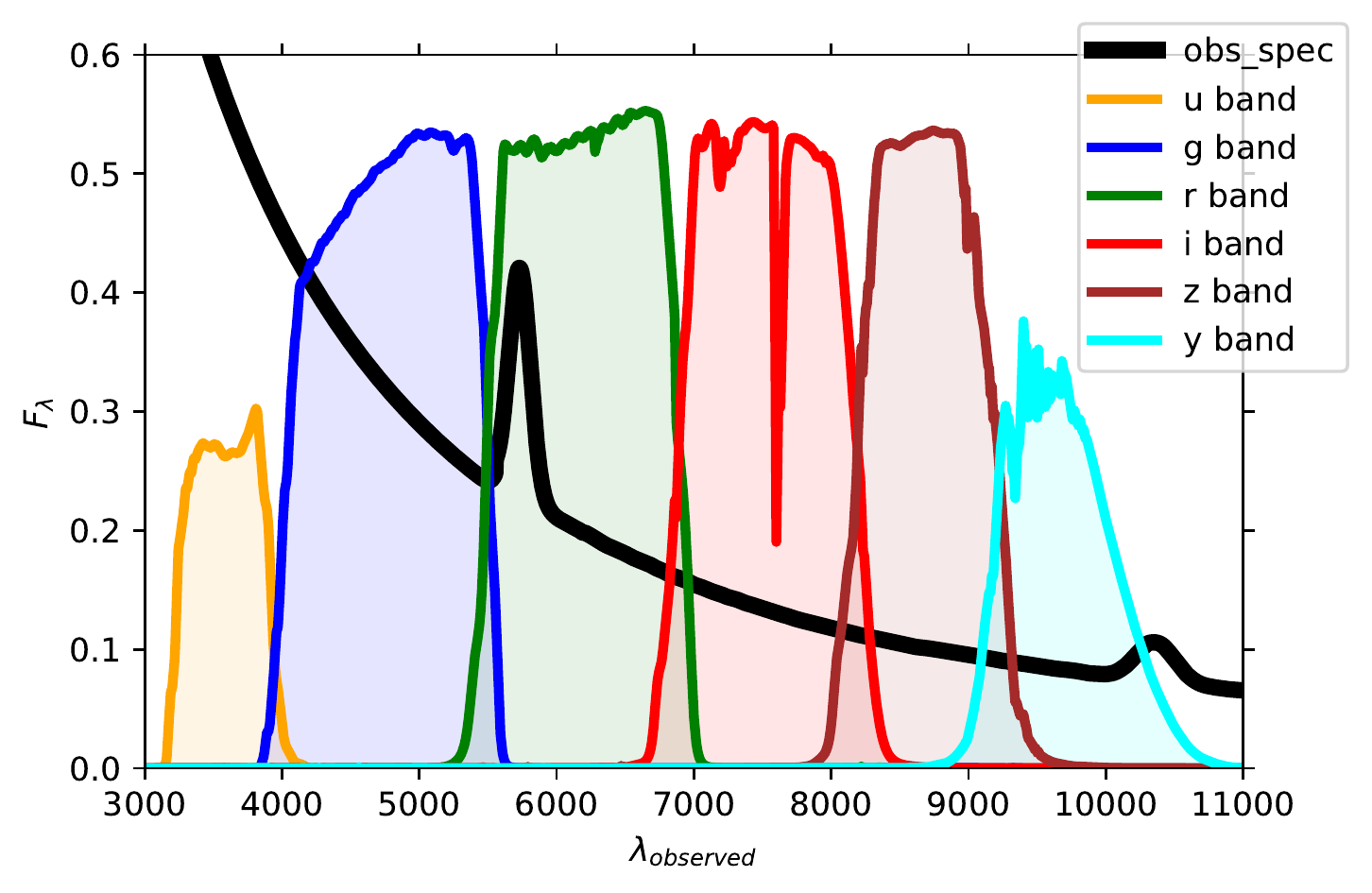}
\caption{Example of the artificial spectrum of an AGN at z = 0 (upper panel) and high redshift z = 2.7 (lower panel). The LSST filters are overplotted.}
\label{fig:widmo}
\end{figure}


We constructed the artificial spectrum of an AGN in the wavelength range of 100 - 2000 nm, which is enough to model the sources with redshift up to 5. The continuum was parameterized with a single slope $\alpha$ ($F_{\nu} \propto \nu ^{\alpha}$, or $F_{\lambda} \propto \lambda^{-2-\alpha}$). We then added the emission lines H$\beta$, MgII, and CIV. The line shapes were parameterized as single Gaussians (MgII was treated as two Gaussian components because it is a doublet), the width was assumed to be the same for all three lines, and it was set by the full width at half maximum (FWHM). Optionally, it can be represented by a Lorentzian shape. The line intensity was set by assuming the line equivalent widths (EWs). As a default, we assumed the EWs characteristic for bright quasars \citep{forster2001}. We also included the broadband contaminants in the form of the FeII pseudo-continuum and the starlight. The FeII continuum was taken from one of the theoretical templates ($d11-m20-20.5-735.dat$) of \citet{Bruhweiler2008}, which fits the spectra of quasars well, for instance, LBQS 2113-4538 ( \citealt{hryniewicz2014}, HE 0435-4312; \citealt{zajacek2021}). Since the model does not contain any kinematic broadening, we applied Gaussian smearing parameterized by the Gaussian width. The FeII pseudo-continuum contributes to the optical and to the UV part of the spectrum.  The spectral shape of the starlight was also taken from the model of \citet{CidFernandes2005}, the version developed by \citet{bruzual2003},  and the parameters were adjusted to fit the bright Seyfert galaxies in \citet{sniegowska2018}. The normalization was the only free parameter of the model.

Next, we calculated the contamination of the three emission lines to all photometric bands for a given redshift and source properties. This was done by folding the spectrum shifted to the observed frame with the profiles of the LSST filters. We first calculated the content of the filter when the selected line was included in the AGN spectrum, then we repeated the calculations setting the EW of the selected line to zero, and next, we calculated the ratio. In this way, we obtained the fractional contamination of each line to each filter, which is a few up to 15\%, depending on the line. Starlight and FeII contributions were always included, so that the contamination was measured against the sum of the continuum and pseudo-continua. This allowed us to confirm the band selection and informed us about the importance of the selected line in the selected band.
The exemplary spectra at two values of the redshift are shown in Figure~\ref{fig:widmo}, and the photometric profiles of the LSST filters are overplotted.

\subsection{Construction of a single dense light curve}
\label{sect:curve_reference}

We first constructed a single dense stochastic light curve. To do this, we used the algorithm developed by \citet{TK1995}.  We usually assumed a broken power-law shape for the underlying power spectrum, with two frequency breaks, $fb_1$ and $fb_2$, and three slopes ($\alpha_1$,$\alpha_2$, and $\alpha_3$). This parameterization is more general than the damped random walk that is frequently used to model AGN light curves \citep{kelly2009}, which would correspond to $fb_1 = fb_2$, and $\alpha_1 = 0$, $\alpha_3 = 2$. 
The random aspect appears when the power spectrum specified in the frequency domain is Fourier-transformed to the time domain because the power spectrum specifies only the value of the Fourier transform, but not the phase. This phase was thus adopted as random, which is justified by studies of AGN light curves. This led to (an almost arbitrary) number of light curves corresponding to the same power spectrum and thus representing the same set of parameters. 
As advised by \citet{uttley2005}, in the next step the exponent of the constructed stochastic light curve was calculated for final use
which allowed us to avoid  negative values when the fluctuations were strong, and it additionally reproduced the standard correlation between the rms and flux and the associated log-normal flux distribution seen in accreting sources. The remaining parameters are the time step in the dense light curve, $\delta T$, and the total duration of the light curve, $T$. The normalization of the curve is provided by the assumed total variance. This light curve later represents the continuum band, which is relatively free from contamination by a strong emission line.

\subsection{Delayed dense light curve}
\label{sect:curve_delayed}

In the next step, we created a delayed dense light curve. This required the assumption of the time delay expected in a given object. In our modeling, we used the radius-luminosity relation derived as 
\begin{equation}
\log{(\tau_{\rm obs} \;[{light-days}])}= \beta + \gamma\log{L_{3000,44}} + \log (1+z),
\label{eq:delay}
\end{equation}
where $L_{3000,44}$ is the monochromatic absolute luminosity at 3000\AA~ in units of $10^{44}$, in erg s$^{-1}$, $\tau_{\rm obs}$ is the time delay in the observer's frame, and $z$ is the source redshift. The values of the coefficients can be taken from the observational studies of the R-L relation \citep[e.g.,][]{kaspi2000,peterson2004,bentz2013}, and they are slightly different for various lines, in particular, the delay for CIV is shorter than the typical delay for H$\beta$ and Mg II \citep[e.g.,][]{lira2018}. This was not included in the simulations. We used the same R-L for all the lines, and the fixed parameters were set as default ($\beta = 1.573$, $\gamma = 0.5$). The choice of the slope was motivated theoretically by the failed radiatively accelerated dusty outflow (FRADO) model of the BLR \citep{czhr2011,czerny2015,czerny2017,naddaf2021,naddaf2022}, which is well justified for H$\beta$ and Mg II, but not for CIV, which belongs to the high-ionization lines and forms in the dustless line-driven wind, closer to the black hole. 

We used the delay given by Equation~\ref{eq:delay} to shift the original dense line {\bf light curve}. Additionally, because the reprocessing in the BLR happens in an extended medium, the original dense curve should be convolved with the response function of the BLR. These response functions were derived observationally for a few nearby sources \citep[e.g.,][]{grier2013,xiao2018,du2018,2021ApJ...907...76H}. Attempts to do this for Mg II are complicated by the presence of the underlying Fe II component \citep{Panda_2021_PhD, prince_CTS_2022}. Therefore, in the current paper, we simply assumed the response function in the form of a symmetric Gaussian shape, with the time shift set by Equation~\ref{eq:delay}, and the width $\sigma_{BLR}$ of 10\% of the same delay. We also performed tests using 
a half-Gaussian shape for this purpose as an exception, as was done in the simulations of the time delay by \citet{vikram2022}.

\subsection{Cadence in all six photometric bands}
\label{sect:cadence}

As argued, for example, by  \citet{malik2022}, general sampling, and in particular, seasonal gaps, is a critical issue for a successful delay recovery. 
Thus, in order to realistically replicate the actual LSST cadence, we used the operation simulator (OpSim) results provided by the VRO-LSST data management team. For the wide-fast-deep (WFD) MS survey, we used the OpSim run baseline\_v2.0\_10yrs and extracted the 5$\sigma$ depth light curves for the six bands (ugrizy), determined for 30-second exposures. We made a random search using a 3.5$^{\circ}\times$3.5$^{\circ}$ search area and limited the sky coordinates (RA, DEC) within 0.01 dispersion. This criterion allowed us to choose roughly the 5$\sigma$ depth for the same (synthetic) source across the six bands. We made a similar search for the deep-drilling field surveys, where we use the OpSim Run: ddf\_v1.7\_nodither\_10yrs. We used these two cadences in most of our simulations, referring to them as DDF and main survey (MS) for simplicity. The 5$\sigma$ depth in principle informs us about the photometric accuracy of the measurement, depending on the adopted luminosity and redshift of the source, but this is not yet incorporated in the software, and we used a fixed photometric accuracy.  However, the typical limit in the g band in the selected field is 24.5 mag, which corresponds to a 5$\sigma$ detection of an AGN with $\log L_{3000} = 43.814$ in erg s$^{-1}$ ( we use values of the luminosity L in units of erg s$^{-1}$ throughout the paper), according to the online AGN calculator \citep{kozlowski2015}. This means that a quasar with an adopted $\log L_{3000} = 44.7$ at redshift 2 will be detected with 0.06 mag error and a quasar at $\log L_{3000} = 45.7$ will be detected roughly with 0.02 mag error. However, some of the exposures are repeated two to three times within 6 hours for the MS and five to ten times in DDF within a very short time period of 5 to 10 minutes. While these multiple observations do not sample the AGN variability in practice, they effectively lower the error. As a default, we used an even much lower error to emphasize the problems that are directly caused by the red-noise character of the light curves combined with the planned sampling. 

We selected two bands for the time-delay measurement: One band that is strongly contaminated by one of the broad emission lines, and the other band was free of contamination, which closely represents the continuum, and neighbors the selected contaminated band. Since in the future we may wish to also use the photometry from other bands to model the continuum, we currently read all the simulated observational dates from the LSST cadence simulator for a selected specific position on the sky and a specific cadence model. This is currently done externally; the cadence is extracted using the simulated databases from the LSST operation simulator, which are processed locally using {\sc python} and {\sc SQLite} and stored in the form of an ASCII file. 


\subsection{Creating two simulated photometric light curves}

With two dense light curves representing the continuum and the line emission, as described in Sections~\ref{sect:curve_reference} and \ref{sect:curve_delayed}, as well as simulated dates of the measurements in the two selected bands (Section~\ref{sect:cadence}), we now construct the modeled light curves. They were constructed by adding the reference curve and the delayed curve, but with the delay curve normalized by the level of line contamination, as specified in Section~\ref{sect:contamination}, and by interpolation to the planned cadence. Observations in the two bands are not simultaneous, they simply follow the set LSST cadence for the chosen location in the sky. Only one of the two constructed curves is strongly contaminated by the BLR, as designed, so that it contains the relatively delayed signal, typically of about a few percent, depending on the redshift and adopted strength of the lines. 

At this stage, we also included the additional noise due to the expected photometric error ($P$) in magnitudes, which (for the small error) is equivalent to the dimensionless fractional error. This was done assuming the photometric accuracy in magnitudes and by adding a Poisson noise to the curve by multiplying each data point flux by $(1 + P \sigma_P)$, where $\sigma_P$ is the random number representing the Gaussian distribution with zero mean and a dispersion of 1.

\begin{figure}  
\centering
\includegraphics[width=\columnwidth]{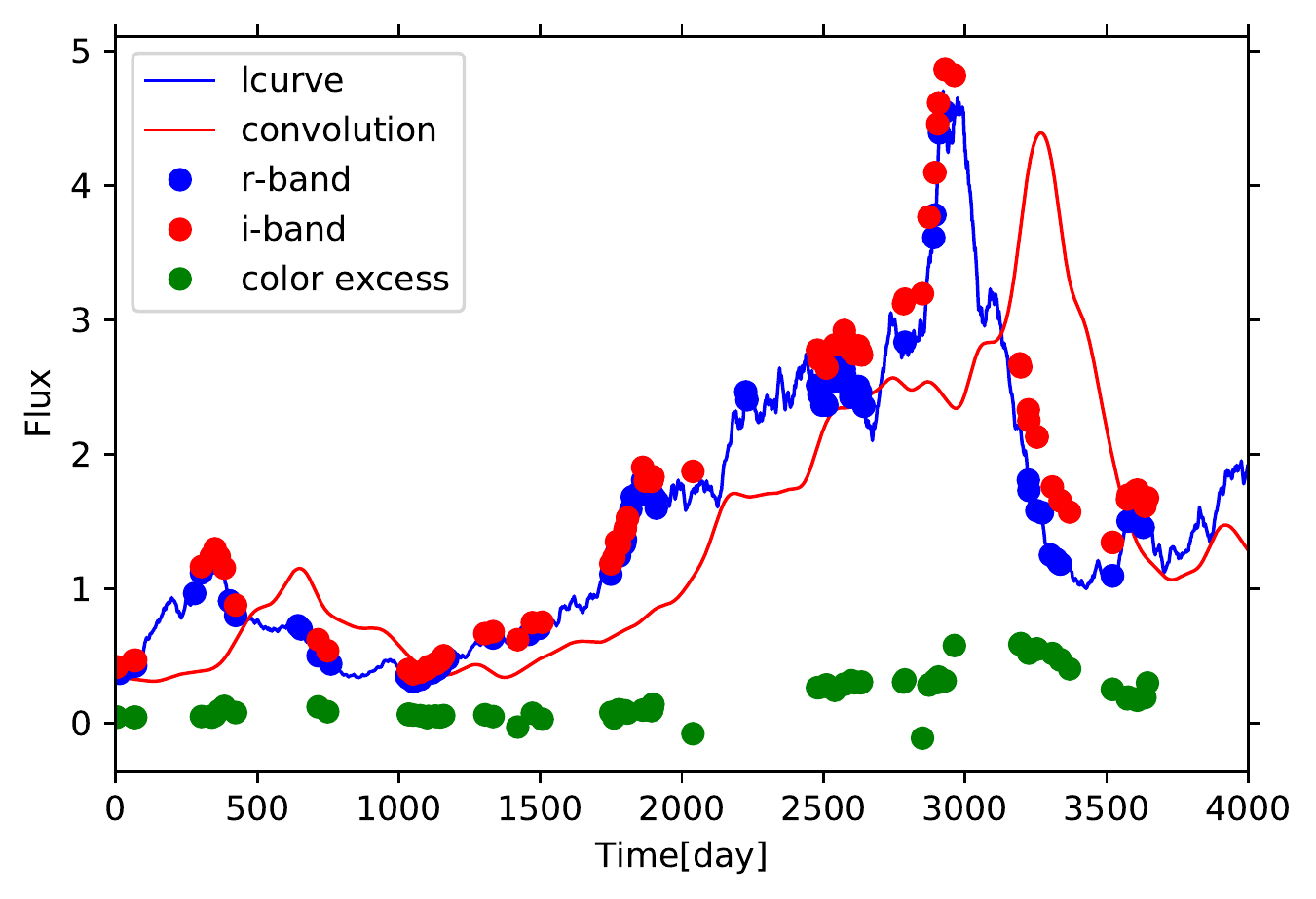}
\caption{Example of the artificial dense light curve (continuous blue line), its convolution with the BLR response (continuous red line), observational points in $i$ band set by the cadence (red circles), and observational points in $r$ set by the cadence, with contamination from the CIV line (blue circles). Green points represent the net contamination for $\epsilon_i = 1.0$ (see Equation~\ref{eq:subtraction}). The delay is calculated between the green and blue points. We adopt standard values of the parameters from Table~\ref{tab:parameters} and $z = 2.7$.} 
\label{fig:curves}
\end{figure}

In this approach, we neglected the time delays between the nearby continuum bands used in the process, which are generated by the reverberation in the accretion disk. This is an approximation made for simplicity because we concentrated on the plausibility of the recovery of the line delays. The intrinsic continuum delays between the continuum bands are much shorter, about one day to a few days, and their modeling involves the assumption of the height of the irradiating source \citep[see, e.g.,][for the time-delay plots]{2021kammoun}. The measured delays frequently appear to be longer, but this is quite likely a result of the BLR contamination \citep[e.g.,][]{netzer2022}. Sparse monitoring of the main survey should not be affected by intrinsic continuum time delays. DDF monitoring might be used to disentangle the intrinsic accretion disk delays and the BLR, but we do not address this issue in the current paper. It is important to note that the BLR time delays basically scale with the monochromatic luminosity as the square root (see Equation~\ref{eq:delay}, where $\gamma \sim  0.5$), and the same scaling is expected from the accretion disk reverberation \citep[e.g.,][]{collier1999,cackett2007} so that the intrinsic continuum time delay should always be shorter for all objects. 
An example of the two dense light curves representing the uncontaminated photometric channel, the convolution representing the contaminated channel, and the observational points representing the actual cadence is shown in Figure~\ref{fig:curves}.

\subsection{First stage of the preparation for the time-delay measurement}
\label{sect:delay_recalib}

We initially tested whether the time delay could be directly measured from the two photometric curves. However, the measurements were very inconclusive because the second light curve contained only a few percent of the delayed line emission. As discussed in previous studies of photometric reverberation mapping \citep{pozo2012,pozo2015}, the varying AGN continuum must be removed before cross-correlation techniques are used. This can be achieved by subtracting a fraction of the continuum traced by a band with negligible line contribution. To improve the chances of the delay measurement, we therefore first subtracted the relatively uncontaminated curve $F1(t)$ from the contaminated curve, $F2(t)$,
\begin{equation}
\label{eq:subtraction}
F22_i(t) = F2(t) - \epsilon_i F1(t),~~ i=1, \ldots, 10
,\end{equation}
to select ten values of the coefficient, $\epsilon_i,$ equally spaced between 0.85 and 1.15. This required interpolation because $F1$ and $F2$ are not measured at the same moment. An example of this subtraction is shown in Figure~\ref{fig:curves} with the green points. The time delay is measured for all ten values of $\epsilon_i$, and the time delay was later set to the value that favored the quality of the time-delay fit.

\subsection{Time-delay measurements}
\label{sect:delay}

Finally, we determined the time delay using one of the two methods. The default method used in the current paper is the $\chi^2$ method (for details, see \citealt{czerny2013} and \citealt{bao2022}). Optionally, the interpolated cross-correlation function (ICCF) can be used, which is described in detail by \citet{Gaskell1987} and \citet{peterson1998,peterson2004}. We searched for the delay imposing the lower time-delay limit at 0.25 of the assumed delay, and the maximum time delay was determined as the lower of the two values: half of the total duration of the campaign, and 1.9 of the assumed delay. The errors in the delay measurement in both cases were set by the dispersion (i.e., standard deviation) in the delay measurements obtained from 100 statistical realizations of the initial stochastic light curve described in Section~\ref{sect:curve_reference}.

To measure the time-lags, we also tested the Javelin code (\citealt{2011ApJ...735...80Z,2013ApJ...765..106Z}). We ran numerous tests on the simulated data. The method, when properly applied, takes more time and does not give better results. We discuss this issue in Appendix A.  

\subsection{Assigning representative parameter values}

\begin{figure*}[!htb]  
\includegraphics[width=0.245\textwidth]{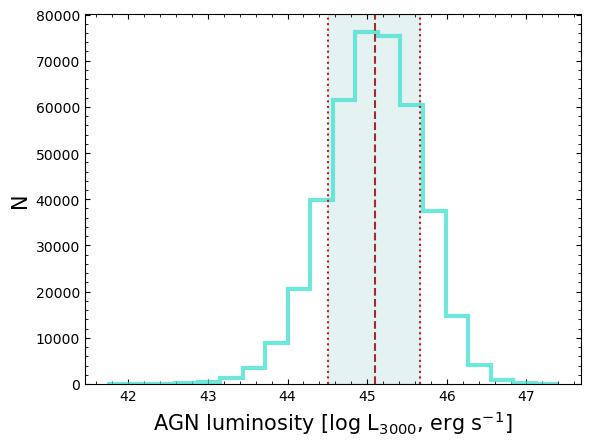}
\includegraphics[width=0.245\textwidth]{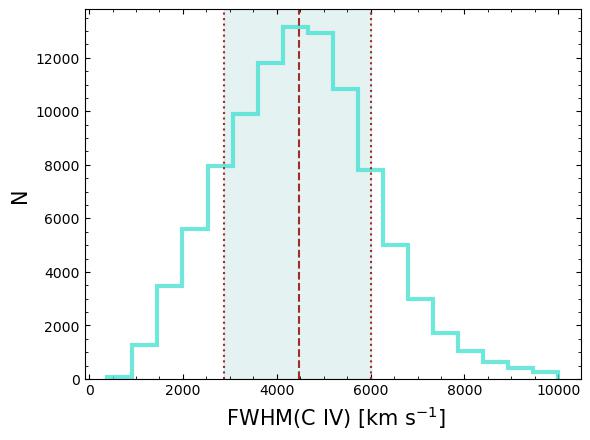}
\includegraphics[width=0.245\textwidth]{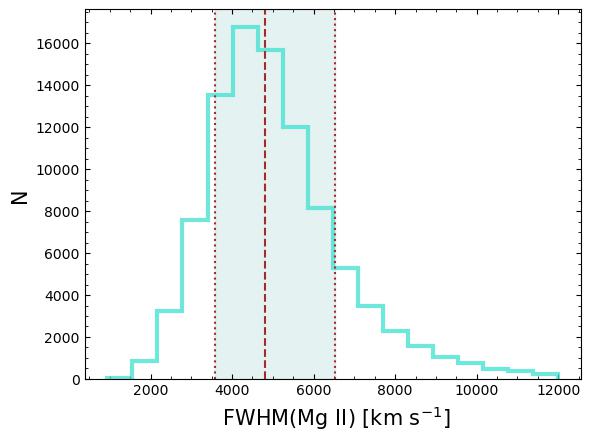}
\includegraphics[width=0.245\textwidth]{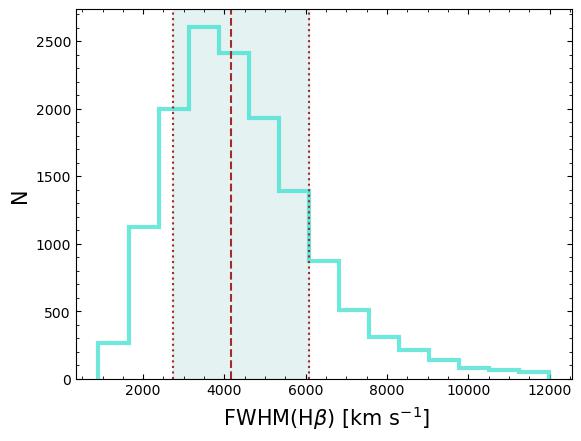}

\includegraphics[width=0.245\textwidth]{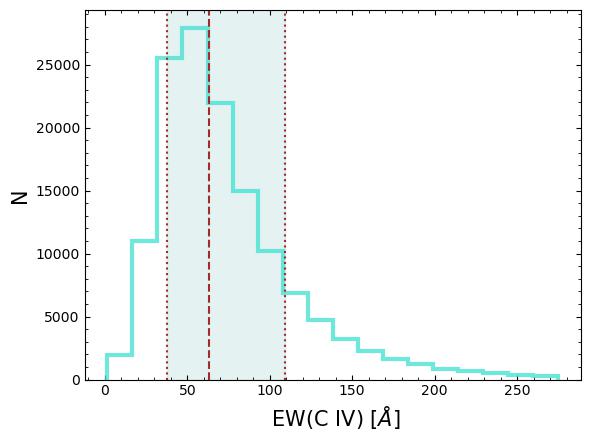}
\includegraphics[width=0.245\textwidth]{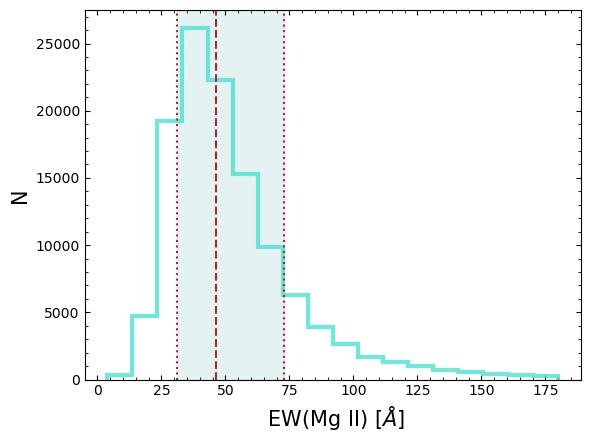}
\includegraphics[width=0.245\textwidth]{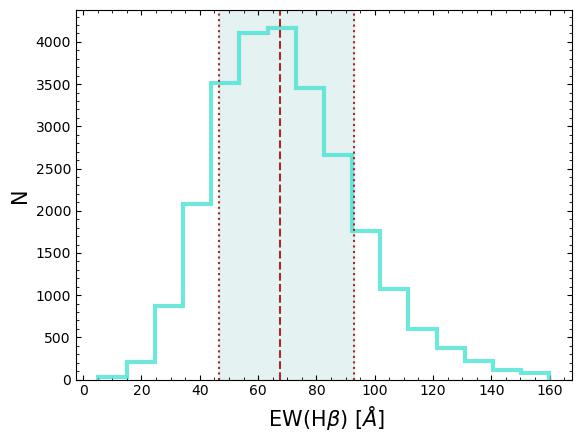}
\includegraphics[width=0.245\textwidth]{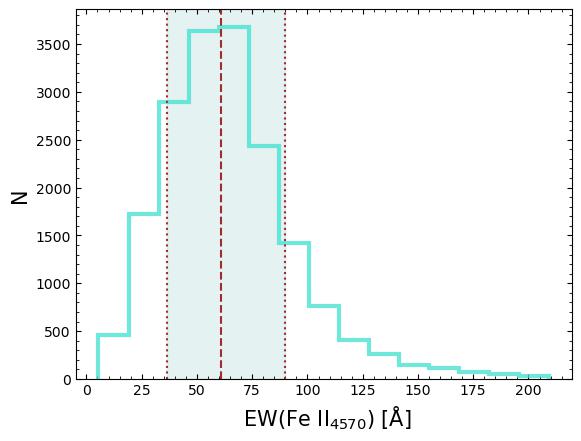}
\caption{Distribution of the quasar luminosity $L_{3000\AA}$, line widths (FWHM of C {\sc iv}, Mg {\sc ii} and H$\beta$) and equivalent widths (C {\sc iv}, Mg {\sc ii}, H$\beta$ and optical Fe {\sc ii}) from the DR14 quasar catalog \citep{sdss-dr14-qso-catalog}. For each distribution, we show the median (dashed lines) and the 16th$^{\rm }$ and 84th$^{\rm }$ percentiles (dotted lines). These statistics are reported in Table \ref{tab:table-sdss_dr14} for each of these parameters.}
\label{fig:distributions}
\end{figure*}

Since the model has many parameters, we first set the representative parameters that were finally included in Table~\ref{tab:parameters} to show the basic trend more easily. We used the SDSS DR14 QSO catalog \citep{sdss-dr14-qso-catalog} to obtain the distribution of the quasar luminosities, line intensities, and line widths. The parent sample from \citet{sdss-dr14-qso-catalog} contains spectroscopically measured parameters such as line and continuum luminosities, line widths and equivalent widths for 526\,265 SDSS quasars. We considered the distribution of the relevant parameters that serve as input to our code, that is, the continuum luminosity at 3000\AA\ (or L$_{\rm 3000\AA}$), the line widths (FWHM) for the prominent broad emission lines (C {\sc iv}, Mg {\sc ii,} and H$\beta$), and their equivalent widths (EWs). We also considered the EW for the optical Fe {\sc ii} emission integrated between 4434-4684\AA,\  which is an important contaminant and coolant in the BLR \citep{bg1992, verner_etal_1999, shen_ho_2014, marinello_etal_2016, panda_etal_2018, marziani_etal_2018, panda_2022}. For the L$_{\rm 3000\AA}$, the DR14 QSO catalog provides a quality
flag to assess the goodness of fit in addition to the luminosity and corresponding error for each source. the quality flag = 0 corresponds to a good-quality measurement, while measurements with the quality flag $>$ 0 may not be reliable either due to poor S/N or poor spectral decomposition. We therefore filtered sources with L$_{\rm 3000\AA}$ $>$ 0 and a corresponding quality flag = 0. This returned 405\,077 sources. The first panel in Figure \ref{fig:distributions} shows the distribution of L$_{\rm 3000\AA}$ for these sources. We similarly created subsamples for the FWHMs and EWs of the broad emission lines. Their distributions are reported in the other panels of Figure \ref{fig:distributions}. For the FWHMs and the EWs, there are no quality flags. In addition to filtering for sources with \textit{\textup{value}} $>$ 0 (here \textit{\textup{value}} represents the FWHM or EW for the emission lines of interest), we therefore employed an additional filter: \textit{e$_{\rm value}$}/\textup{value} $<$ 0.1 (where \textit{e$_{\rm value}$} represents the errors for the corresponding \textit{\textup{value}}). We realized that these original distributions had a tail with absurdly high values of about 4-5$\times 10^5$ for the FWHMs and $\sim$10$^8$ for the EWs. To filter these erroneously fitted cases, we further restricted each of our subsamples within an upper limit $\gtrsim$99th$^{\rm }$ percentile. This upper limit was employed uniquely for each case. The final distributions thus obtained are shown in the remaining panels in Figure \ref{fig:distributions}. The overall counts in each subsample, the median value, and the respective 16th$^{\rm }$ and 84th$^{\rm }$ percentiles for each distribution are tabulated in Table \ref{tab:table-sdss_dr14}. We note that this final filtering to restrict the upper limits of the subsamples has no noticeable effect on the median 16th$^{\rm }$ and 84th$^{\rm }$ percentiles per distribution. The plots from this catalog are shown in Figure~\ref{fig:distributions}. 
The two exemplary values of the $ \log L_{3000}$ luminosity used later in most simulations roughly correspond to 1 $\sigma$ deviation from the mean, that is, 16 \% of quasars are expected to be brighter than $\sim 45.7$, and 84 \% are brighter than $\sim 44.7$.

The line width $\sim 4000$ km s$^{-1}$ is quite representative for all lines. Line equivalent widths are about 60 \AA~ in the studied sample. 
We compared this with the distribution of line equivalent widths from the Large Bright Quasar Survey \citep{forster2001}. This survey reports the EWs of 105 \AA~for H$\beta$, 52.7 \AA~for CIV, and 35.6 \AA~for the Mg II narrow component, and their broad component most likely (partially or mostly) represents the Fe II contamination. The strength of H$\beta$ is then much higher, and we used a higher value as a default value in our simulations, but we later testes the sensitivity of the results to the adopted parameters.

The representative LSST quasars will not necessarily have the same statistical properties because  theLSST will reach considerably deeper \citep{Science_Book_2009,LSST_Ivezic_etal_2019}. In this study, we did not aim at the use of the luminosity function as was done recently by \citet{shen_X_2020} to make specific predictions. 

\begin{table*}[!htb]
\centering
\caption{SDSS DR14 QSO catalog properties as shown in Figure \ref{fig:distributions}.}
\label{tab:table-sdss_dr14}
\resizebox{\textwidth}{!}{%
\begin{tabular}{l|ccccccccc}
\hline
 & z & \begin{tabular}[c]{@{}c@{}}log L$_{3000}$ \end{tabular} & \begin{tabular}[c]{@{}c@{}}FWHM$_{\rm C IV}$ \end{tabular} & \begin{tabular}[c]{@{}c@{}}FWHM$_{\rm Mg II}$ \end{tabular} & \begin{tabular}[c]{@{}c@{}}FWHM$_{\rm H\beta}$\end{tabular} & \begin{tabular}[c]{@{}c@{}}EW$_{\rm C IV}$ \end{tabular} & \begin{tabular}[c]{@{}c@{}}EW$_{\rm Mg II}$\end{tabular} & \begin{tabular}[c]{@{}c@{}}EW$_{\rm H\beta}$ \end{tabular} & \begin{tabular}[c]{@{}c@{}}EW$_{\rm Fe II}$\end{tabular} \\
 & \multicolumn{1}{l}{} & \multicolumn{1}{l}{[erg s$^{-1}$]} & \multicolumn{1}{l}{[km s$^{-1}$]} & \multicolumn{1}{l}{[km s$^{-1}$]} & \multicolumn{1}{l}{[km s$^{-1}$]} & \multicolumn{1}{l}{[$\rm{\AA}$]} & \multicolumn{1}{l}{[$\rm{\AA}$]} & \multicolumn{1}{l}{[$\rm{\AA}$]} & \multicolumn{1}{l}{[$\rm{\AA}$]} \\ \hline\\
Count & 526265 & 405077 & 96925 & 93398 & 13979 & 136138 & 116981 & 25298 & 18097 \\
16th & 0.944 & 44.51 & 2882.03 & 3576.09 & 2742.90 & 37.86 & 30.95 & 46.68 & 36.54 \\
Med & 1.8327 & 45.10 & 4466.10 & 4800.51 & 4153.97 & 63.26 & 46.18 & 67.35 & 60.96 \\
84th & 2.593 & 45.66 & 6000.93 & 6533.97 & 6084.53 & 109.40 & 72.83 & 92.94 & 89.76 \\ \hline
\end{tabular}%
}
\end{table*}

\subsection{Statistical error of the time-delay recovery}\
\label{sect:stat}

The secondary peaks in the histogram are more problematic (see Appendix). We therefore stress that the reported error bars represent the expected error in a single time-delay measurement if no special tests or a preselection of suitable curves is made, and no special methods sensitive to multiple time delays from a single set of data are employed. There are options that can indeed help to decrease the scatter, and we discuss them extensively in the Appendix. 


\section{Results}
\label{sect:results}

\subsection{Expectation from the main survey}

In this section, we study the prospects of measuring emission line time delays using the data from the main survey, which will cover ten years, but not very densely. We selected the position in the sky. We used the location on the sky centered at (0, -30) for the MS and (9.45,-44.025) for DDF. The sky coordinates (RA, DEC) are reported in degrees. 

We followed the steps described in Section~\ref{sect:method}. To illustrate the modeling further, we selected a source with a standard luminosity (see Table~\ref{tab:parameters}) located at redshift 2.7. In Figure~\ref{fig:curves} we show the two photometric bands, one that is only weakly contaminated (the y band in this case), and the other band, which is strongly contaminated (the r band, in this case, containing the CIV line). We also show the effect of the subtraction described in Section~\ref{sect:contamination}. The two photometric light curves roughly follow each other because the contamination by the delayed line emission is weak, 12 \% in this case. The direct measurement of the delay between the two bands is therefore not effective, but the subtracted curve follows the delayed curve much better, making the delay measurement much more accurate.

\begin{figure} 
\centering
\includegraphics[width=\columnwidth]{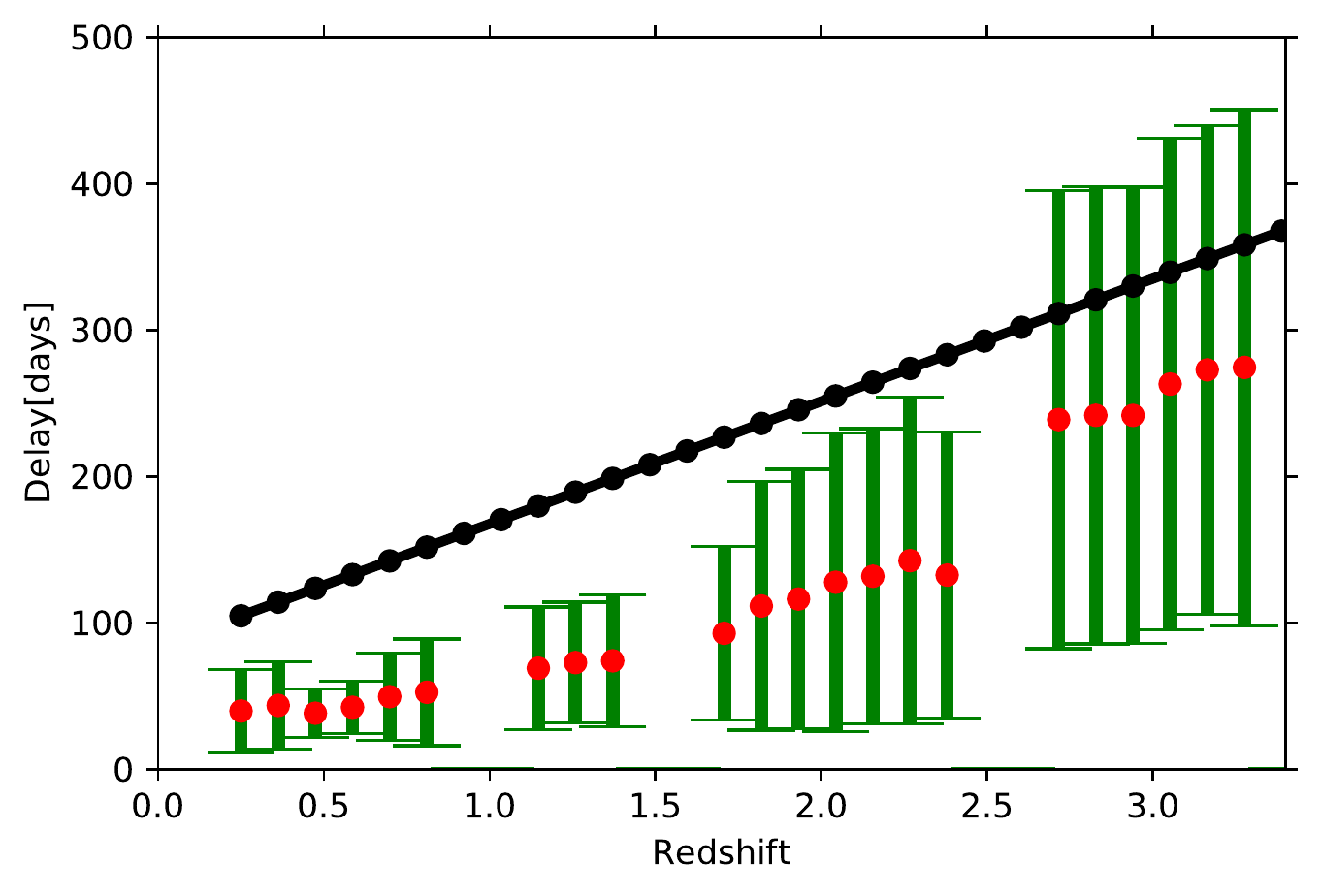}
\includegraphics[width=\columnwidth]{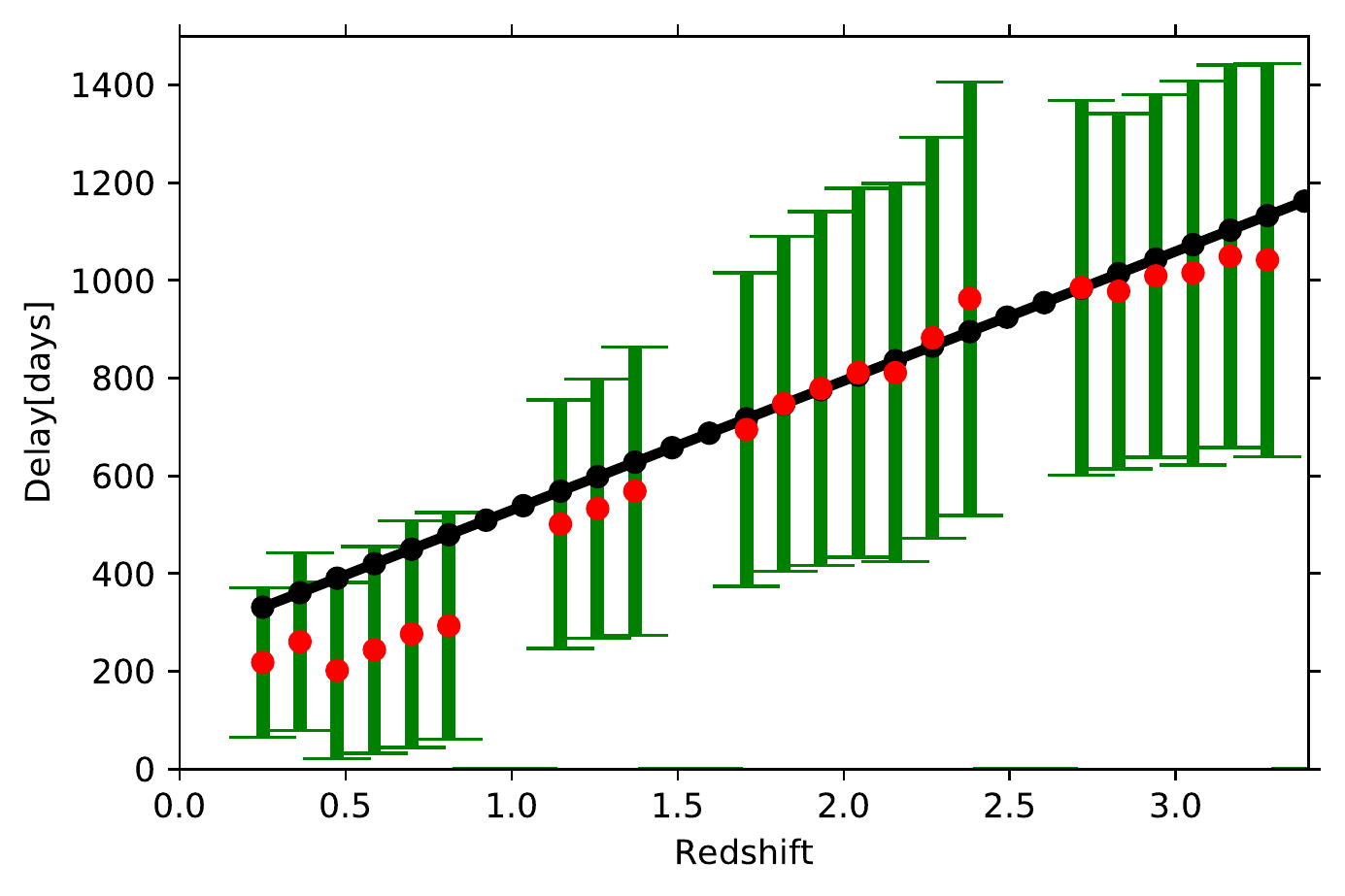}
\caption{Adopted (black points) and mean recovered (red points) time delay as a function of redshift for faint AGN ($\log L_{3000} = 44.7$,  upper panel) and for bright AGN ($\log L_{3000} = 45.7$, lower panel) from the main survey.  Green error bars are the standard deviation that is expected in a single source measurement, as determined by the use of 100 statistically equivalent simulations. The error of the mean recovered delay is 10\% of the dispersion. The other parameters have the standard values given in Table~\ref{tab:parameters}. The redshift gaps correspond to no satisfactory selection of the contaminated and uncontaminated bands.}
\label{fig:MainSurvey}
\end{figure}

Nevertheless, the cadence in the main survey is not quite suitable for time-delay measurements of relatively faint AGN. The upper panel in Figure~\ref{fig:MainSurvey} shows that the measured delay is always considerably shorter than the assumed delay because the timescales corresponding to the actual delay are not well probed. However, for brighter sources, the expected time delay is longer, so that the usual sampling characteristic for the main survey is adequate to recover the line delay at least for $z > 1$ (see the lower panel in Figure~\ref{fig:MainSurvey}). In the actual data analysis, the fainter sources should therefore not be included because the measurement will not be reliable for them. We should also be careful about including the results from too low redshifts, based on H$\beta$. Although bright quasars, with monochromatic luminosities above $\log L_{3000} = 45.7$, in erg s$^{-1}$, are relatively well measured, the expected and recovered time delay at redshifts below 1.0 are still offset.

In our simulations, we adopted the same theoretical time delay for all the lines, even though the CIV line delay is usually shorter. This may cause an additional increase in the error at higher redshifts where the CIV line is used, and in this case, the minimum source luminosity should be even higher.
The delay based on Mg II is overall comparable to H$\beta$ \citep{zajacek2021} so that the error in our approach is smaller.

\subsection{Expectations from the deep drilling fields}

In the DDF area, the number of quasars observed is orders of magnitude smaller than expected from the main survey. However,
the DDF provide a much denser sampling of the light curve, which increases the quality.  This dense coverage importantly allows us to determine some time delays on timescales much shorter than ten years. We thus first discuss the results of the simulations for the entire duration of the project, then for the first two years, and then for the first year of its operation.

\begin{figure} 
\centering
\includegraphics[width=\columnwidth]{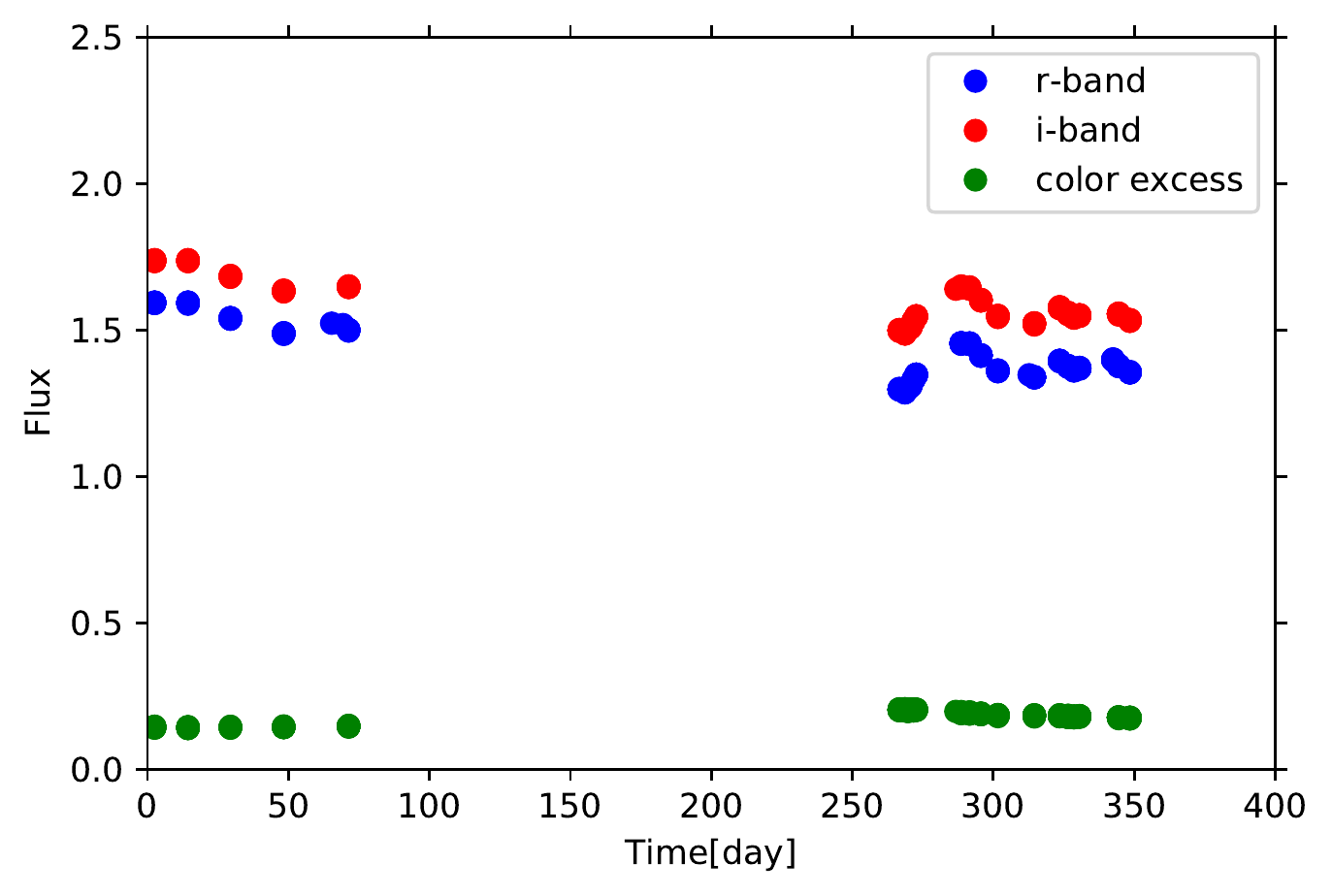}
\includegraphics[width=\columnwidth]{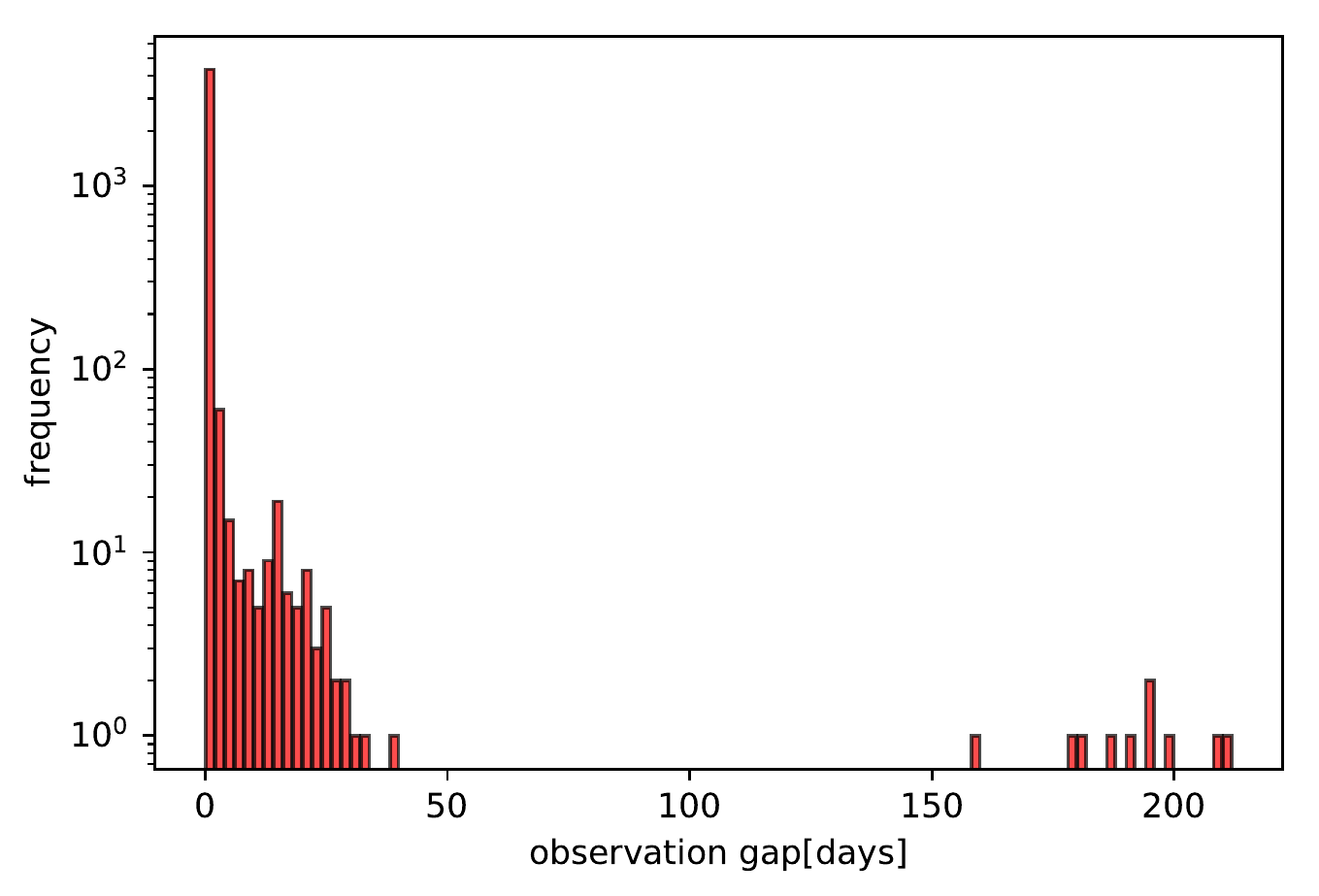}
\includegraphics[width=\columnwidth]{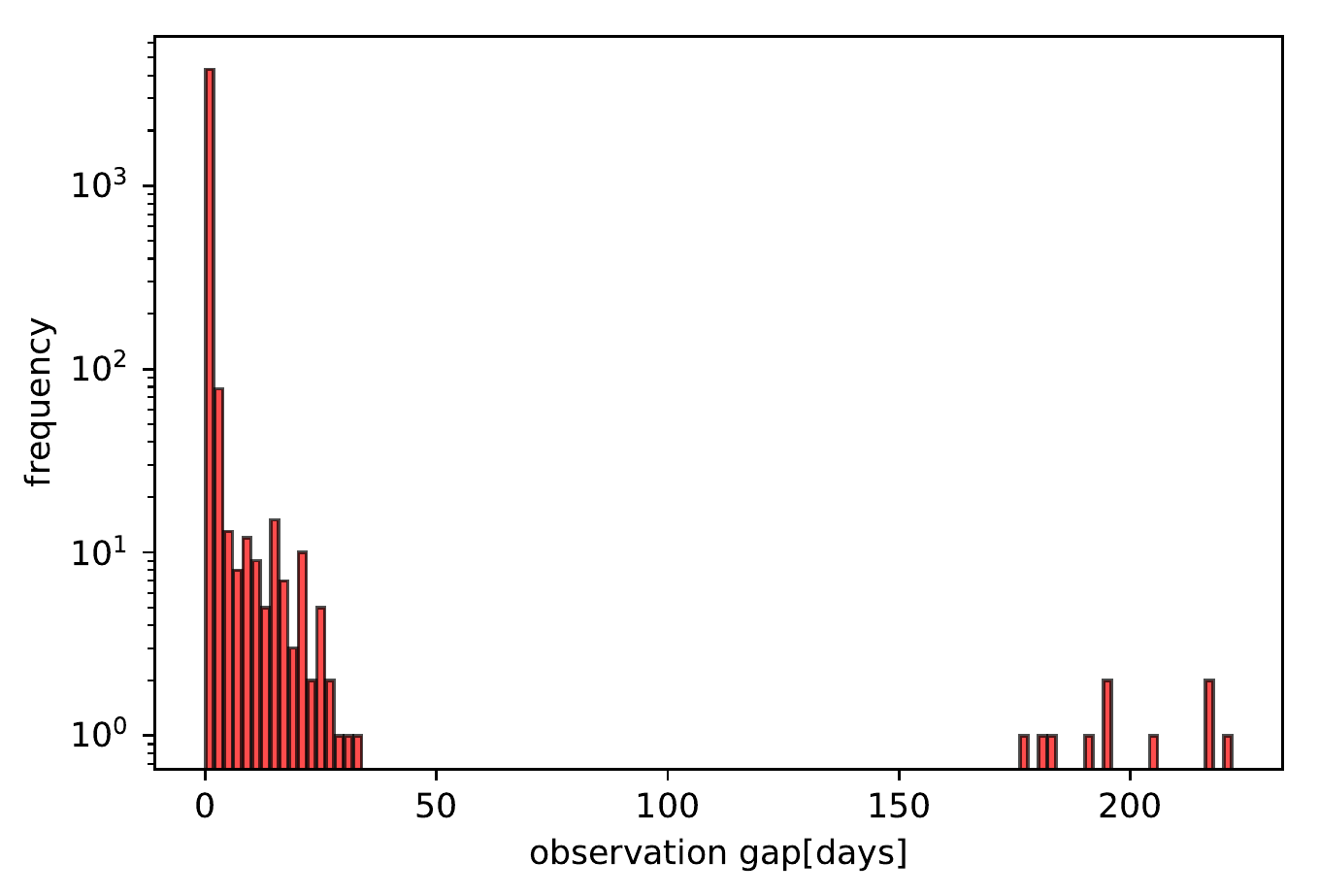}
\caption{Illustration of the DDF cadence issues. Upper panel: Example of the artificial light curve for the first year of observing with DDF cadence in $i$ band (blue circles) and in $r$ band, contaminated by CIV line. Green points represent the net contamination for $\epsilon_i = 1.0$ (see Equation~\ref{eq:subtraction}). The delay is calculated between the green and the blue points. We adopt standard values of parameters from Table~\ref{tab:parameters}, z = 3.276. 
Middle panel: Histogram of the time separation between the consecutive observation dates in $r$ band in the selected DDF field during the whole ten years. Lower panel: Same for the $i$ band.
}
\label{fig:curve_1yr}
\end{figure}

In the specific field that we used in the simulations, the number of observing visits is high: 1056 (u), 2239 (g), 4495 (r), 4496 (i), 2330 (z), and  4436 (y). However, at least in this specific field, 6 visits were typically in the minute time separation that for AGN is equivalent to a single visit, although with an improved S/N. When we count only the visits that are separated by one day or more, then the monitoring is limited to  131 (u), 219 (g), 239 (r),  245 (i), 97 (z), and 226 (y) in ten years, and the time separation is frequently of the order of 2 days with long gaps of about a month (except for the six-month seasonal gaps), which averages to a mean separation of 7 - 9 days in different colors. We illustrate this in Figure~\ref{fig:curve_1yr}, where we plot just the first year of the curve simulated with DDF cadence. Most points are unresolved, and only well-separated point aggregates show up.  This is still much better than the main survey, but not as dense as it might seem from the total number of visits. In our simulation, we used all the visits as they are in the cadence. To illustrate this effect on the whole duration of the survey, we show in the lowest two panels of Fig.~\ref{fig:curve_1yr} the histogram of the time separations in the entire ten-year monitoring. We had to use the logarithmic scale for the vertical axis because the time separations that are shorter than 2 days dominate all other separations by some orders of magnitude. We stress that in the actual computations, no binning was performed. We used the data cadence as it was provided.

\begin{figure}
\centering
\includegraphics[width=\columnwidth]{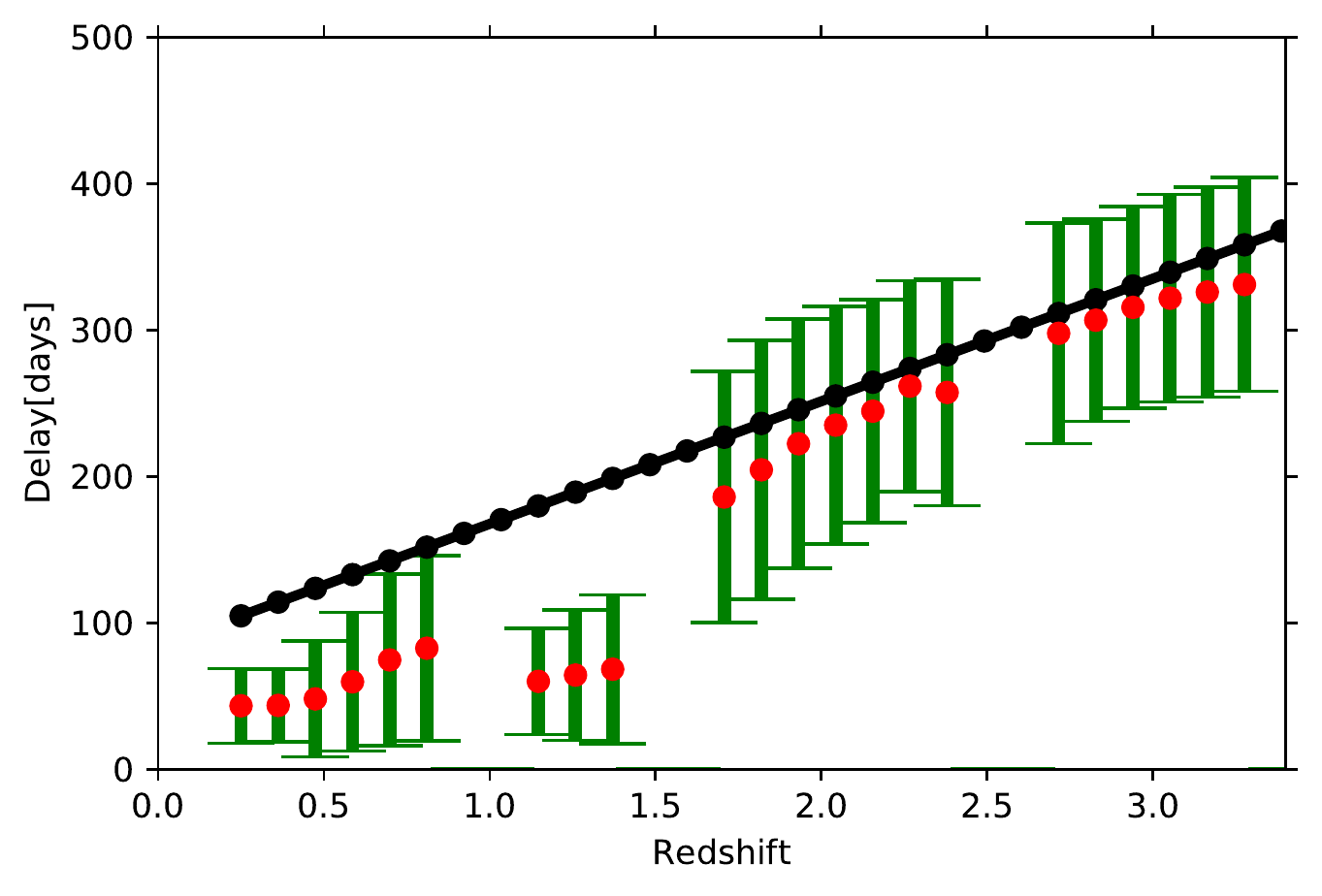}
\includegraphics[width=\columnwidth]{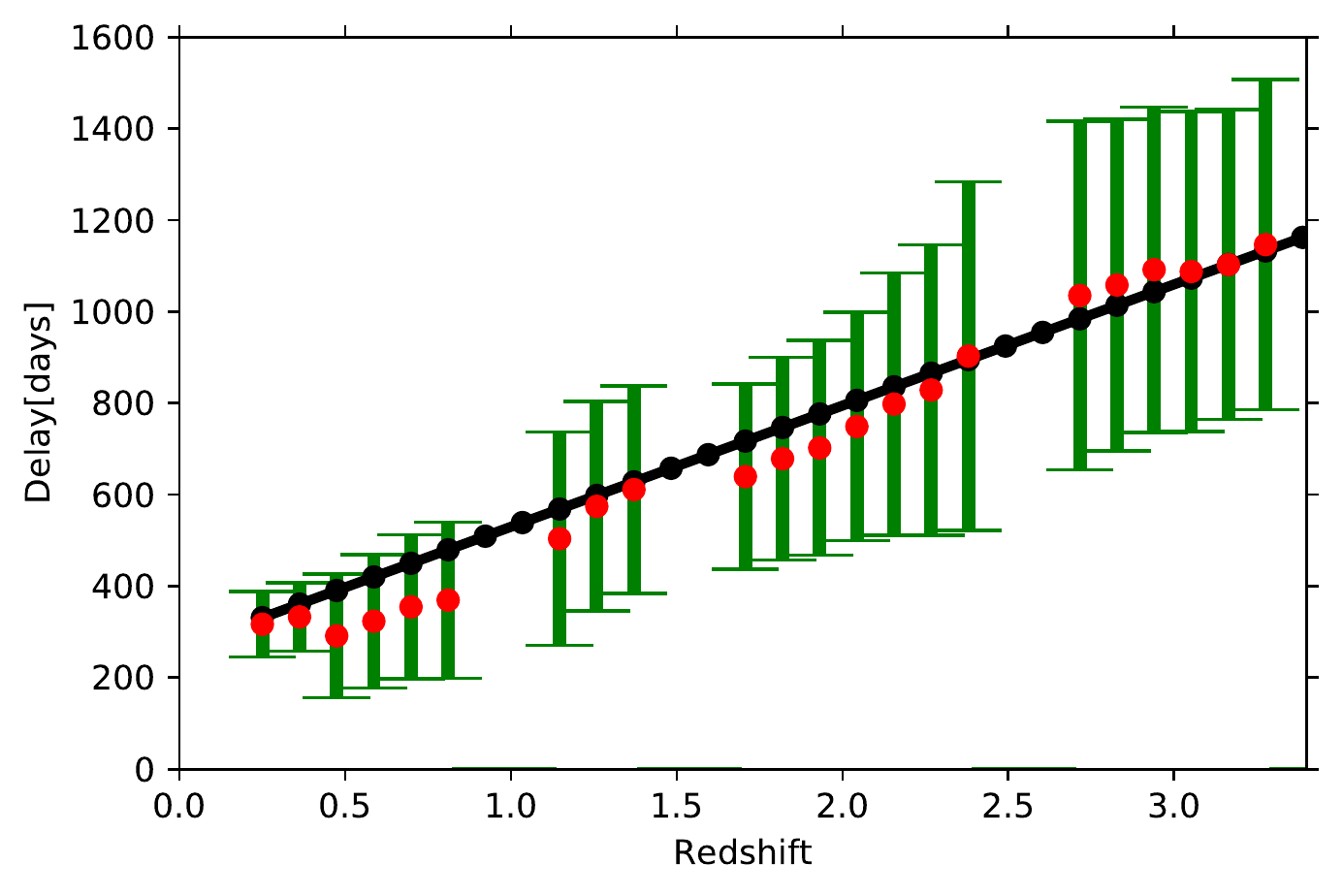}
\caption{Adopted and recovered time delay as a function of redshift for faint AGN ($\log L_{3000} = 44.7$, upper panel) and for bright AGN ($\log L_{3000} = 45.7$,  lower panel) from ten years of observations in the DDF. The other parameters have the standard values given in Table~\ref{tab:parameters}.}
\label{fig:DD10}
\end{figure}

The representative results for the whole ten-year monitoring are shown in Figure~\ref{fig:DD10}. For bright quasars, the results from DDF are not considerably better than from the main survey, except for some improvement at the lowest redshifts (below 1.0), where the denser coverage allows a better determination of the time delay (shorter in this case).

The difference is highly significant for the faint quasars. They are not reliably sampled in the main survey, but those located in the DDF can be well measured at redshifts above 1.8. This is interesting and important because fainter quasars will dominate the quasar population, so that many faint quasars can be detected in the DDF. In the SDSS (see Fig.~\ref{fig:distributions} and Table~\ref{tab:table-sdss_dr14}), about 80\% of the quasars are brighter than  $\log L_{3000} = 44.7$, and in the DDF, they would be well measured, while in the main survey, only about 15\% of the quasars are bright enough to have delay timescales that are long enough to be measured adequately. The low quality at the lowest redshift is partially related to the large gaps between the seasons. The considerable underestimation of the delay for redshifts between 1.0 and 1.5 arises because the expected/assumed time delay in simulations for the adopted luminosity is about 180 days. At redshifts lower than 0.5, the division of the data into separate seasons may help.

\subsection{Expectations from the first year and two years of operation}

\begin{figure} 
\centering
\includegraphics[width=\columnwidth]{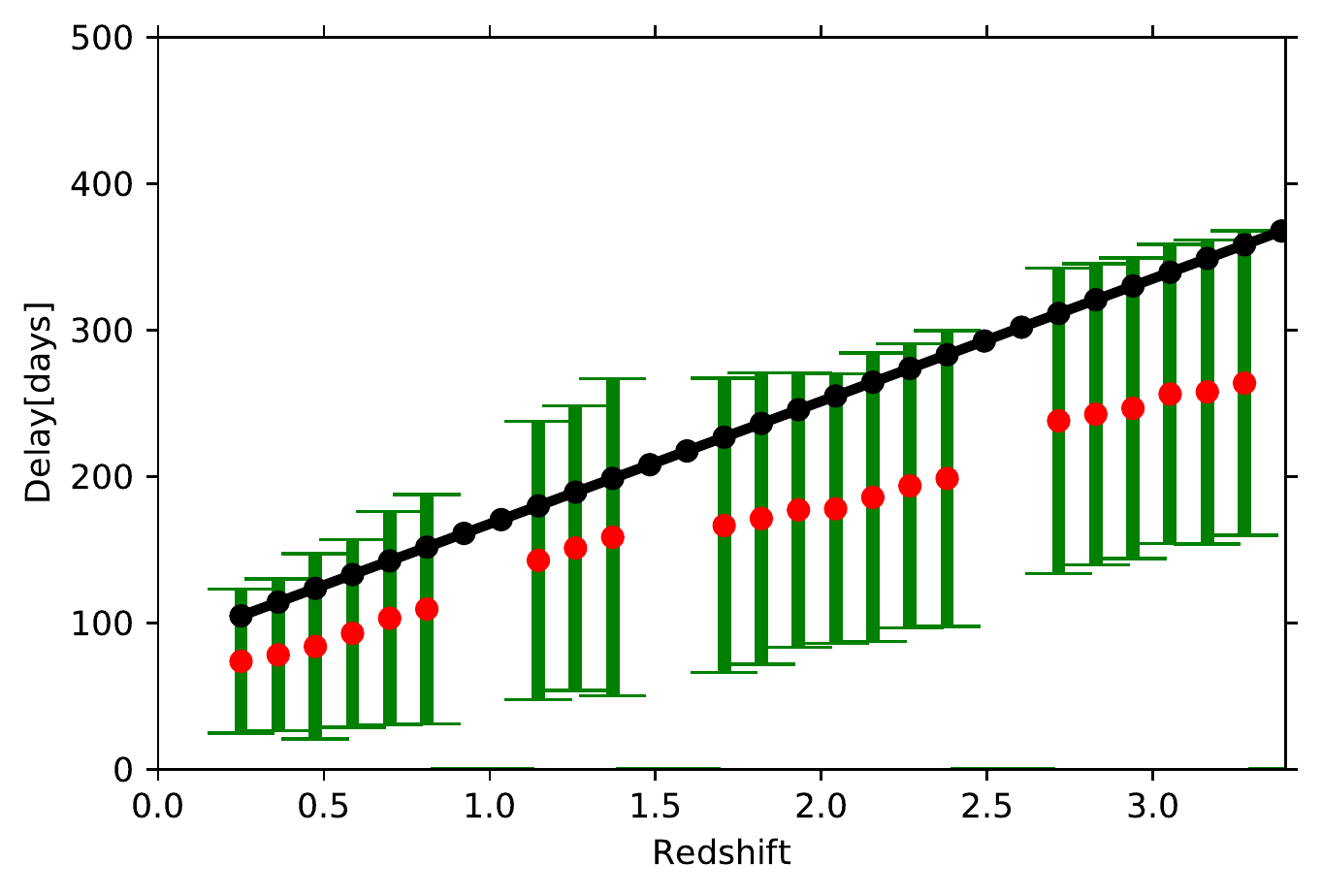}
\includegraphics[width=\columnwidth]{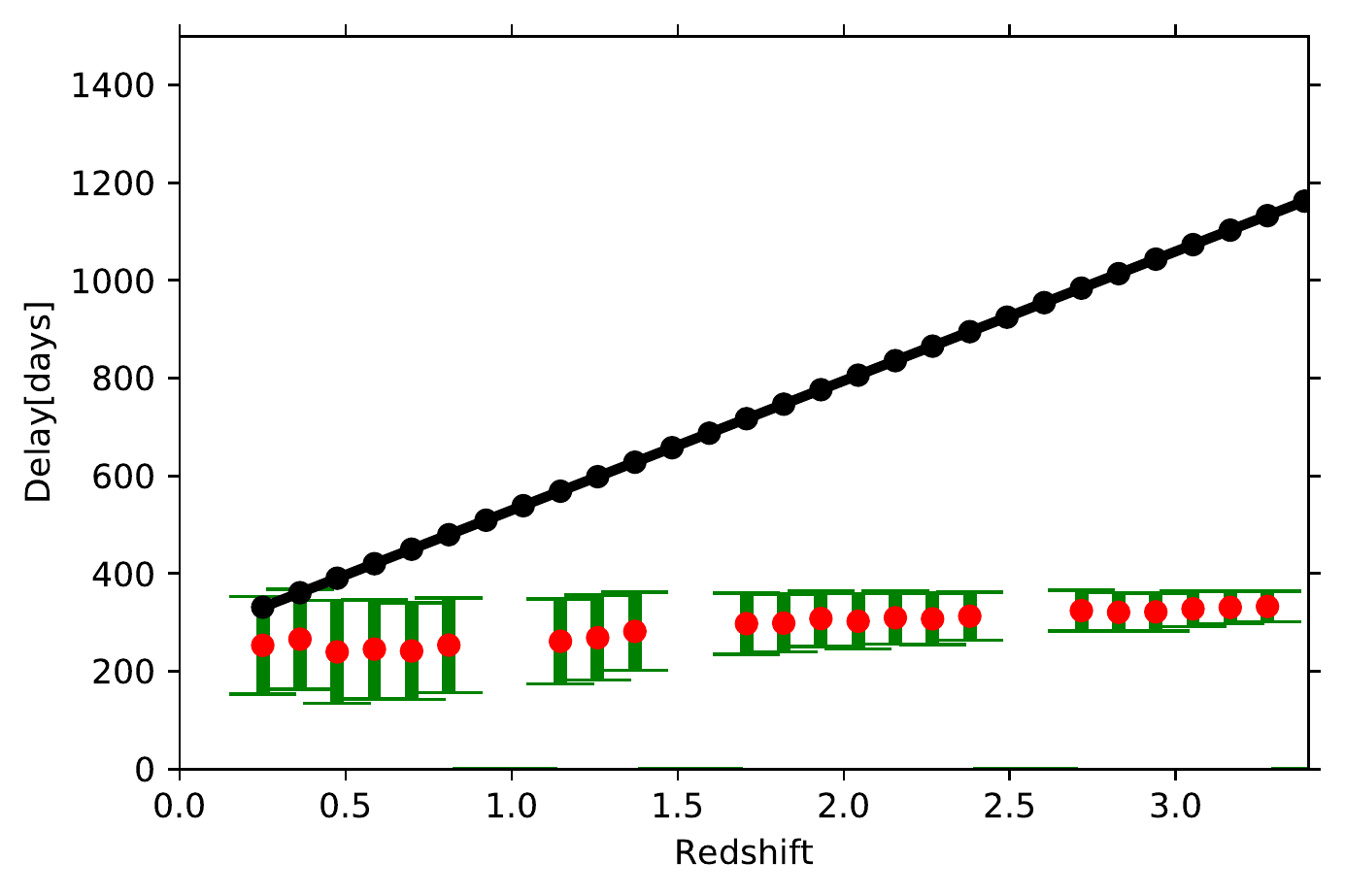}
\caption{Adopted and recovered time delay as a function of redshift for faint AGN ($\log L_{3000} = 44.7$, upper panel) and for bright AGN ($\log L_{3000} = 45.7$, lower panel) from two years of observations in the DDF. The other parameters have the standard values given in Table~\ref{tab:parameters}.}
\label{fig:DD2}
\end{figure}

We first started with a more conservative approach and analyzed the possibility of obtaining interesting results from two years of data. We do not expect any reliable results from the main survey, taking into account the available sampling. However, the cadence of the DDF is much denser, so that time delays of about a year might be measured.
In the case of faint AGN, the expected delays are shorter than 400 days, so that the measurement is possible (see Figure~\ref{fig:DD2}, upper panel). The measurements and the predictions are within 1$\sigma$ error for all redshifts. The mean values are systematically offest from the expected values for all redshifts, however. This offset is caused by data sets that are too short. The measurements are clearly useful for some statistical studies, although a systematic offset of $\sim 40$ \% should be included. This offset will depend on the exact source luminosity, so that appropriate accompanying simulations would be necessary to improve the quality of the results.

In the case of bright AGN, the expected time delays are so long that only the results for objects at redshifts lower than 0.3 are potentially useful when only two years of data are available. Otherwise, the recovered delay saturates at the maximum allowed by the code, which is set at a minimum between half of the duration of the data set and twice the expected time delay (see Figure~\ref{fig:DD2}, lower panel).

\begin{figure} 
\centering
\includegraphics[width=\columnwidth]{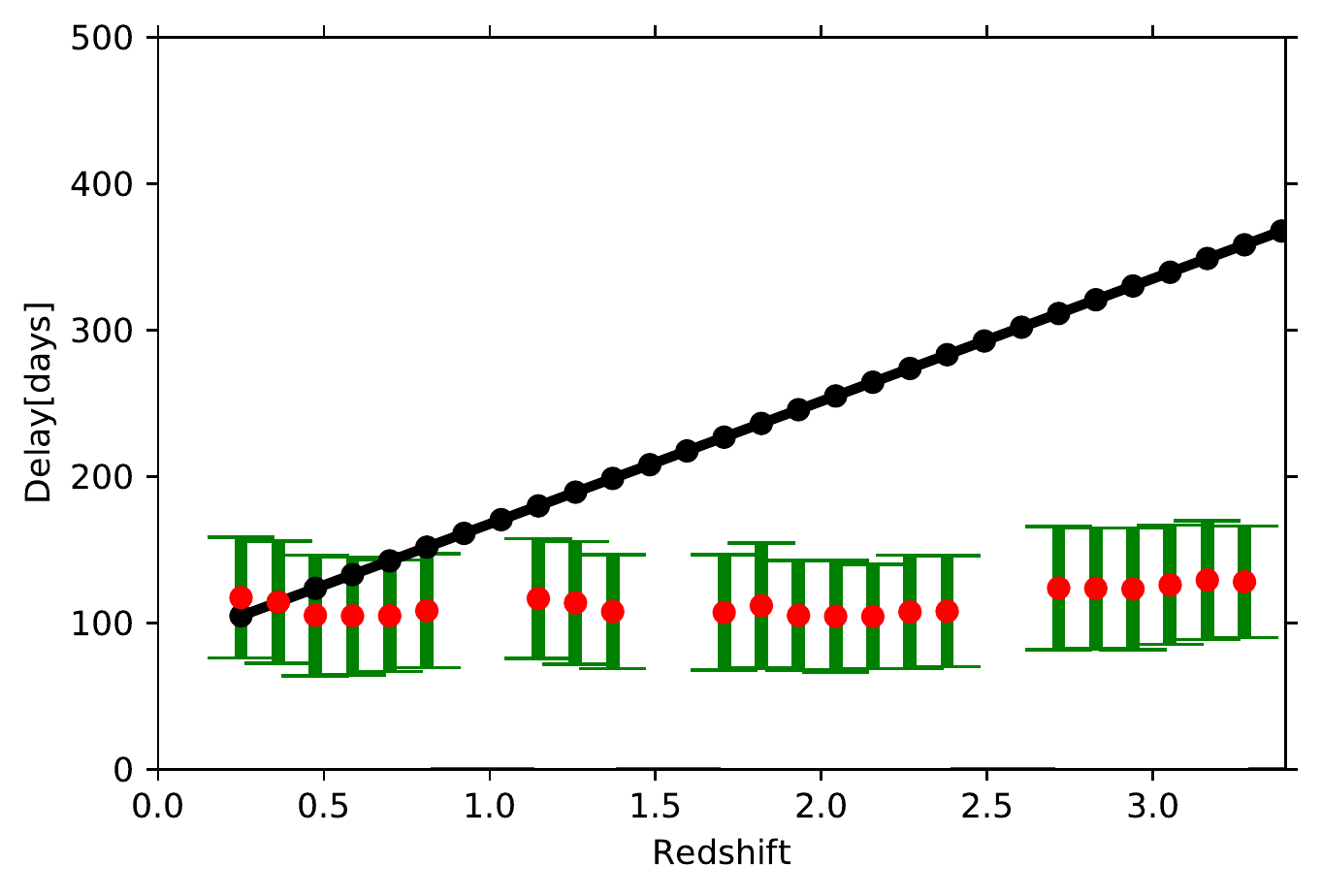}
\includegraphics[width=\columnwidth]{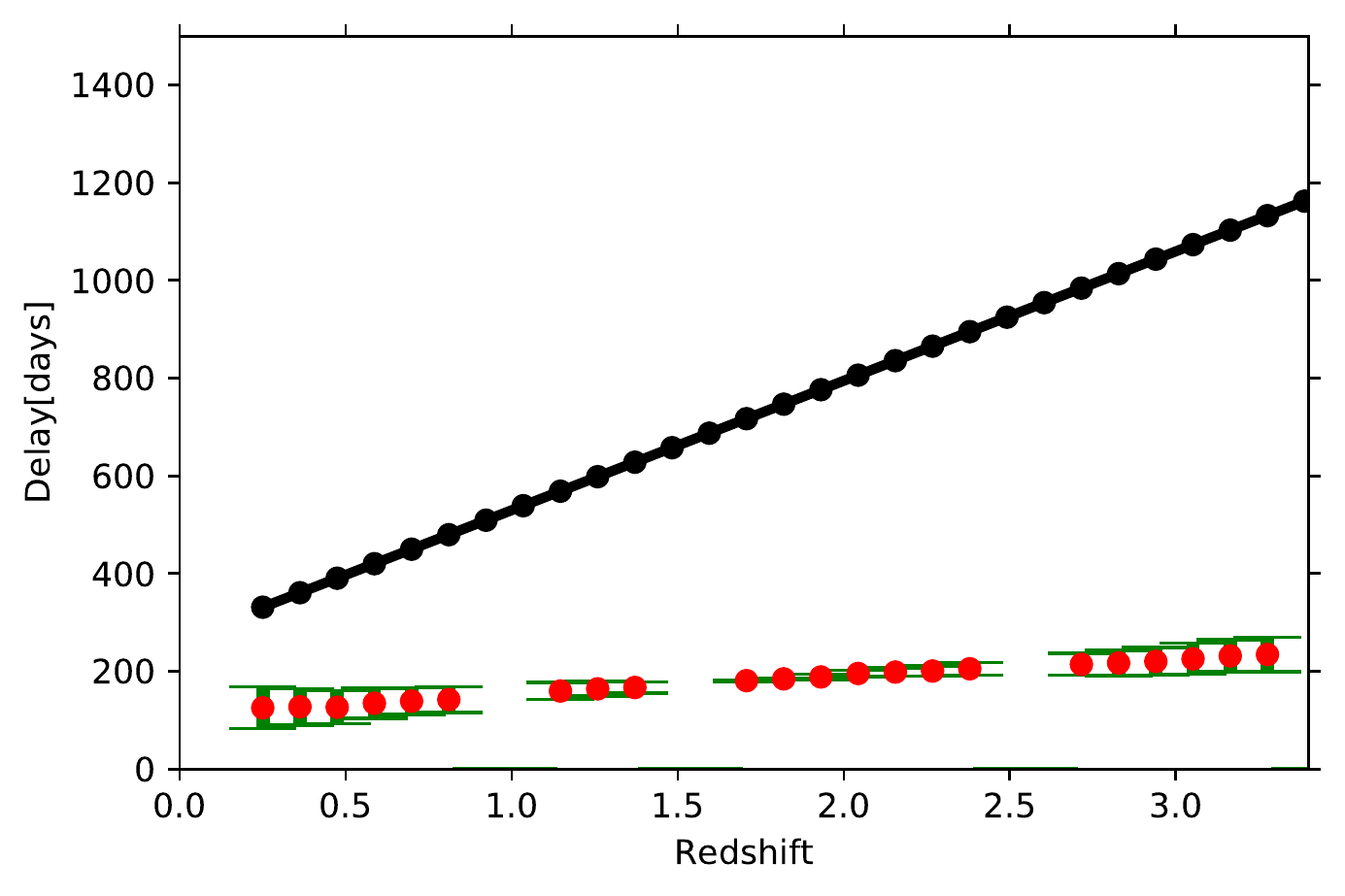}
\caption{Adopted and recovered time delay as a function of redshift for faint AGN ($\log L_{3000} = 44.7$, upper panel) and for bright AGN ($\log L_{3000} = 45.7$, lower panel) from the first year of observations in the DDF. The other parameters have the standard values given in Table~\ref{tab:parameters}.}
\label{fig:DD1}
\end{figure}

If only the first year of the data is used, the situation is even more difficult. For bright sources, the measurements are unreliable for any redshift, and in the case of faint sources, only the results from low redshifts are promising (see Figure~\ref{fig:DD1}). In any case, AGN at a redshift higher than 0.7 are beyond reach, and conservative searches should rather use an even lower redshift limit of $\sim 0.4$. Nevertheless, it is encouraging overall that some AGN emission-line time delays can be measured from such a short monitoring, with such a highly nonuniform cadence.

\subsection{Prediction sensitivity of the adopted parameters}
\label{sect:varia}

Since the number of parameters in our model is large and the representative parameter choice is justified, we tested the dependence of the results only for a selection of assumed parameters. Some of the parameters, such as the FWHM, are of no importance except for a small change in the photometric bands available for time delay measurements. 

We tested the dependence on the parameters by mostly concentrating on the first year of the LSST data, since these data will be available relatively soon and can be used to draw important conclusions as soon as possible. However, the trend generally applies to the full ten-year DDF survey as well as to the main survey mode.

\begin{figure}
\centering
\includegraphics[width=0.8\columnwidth]{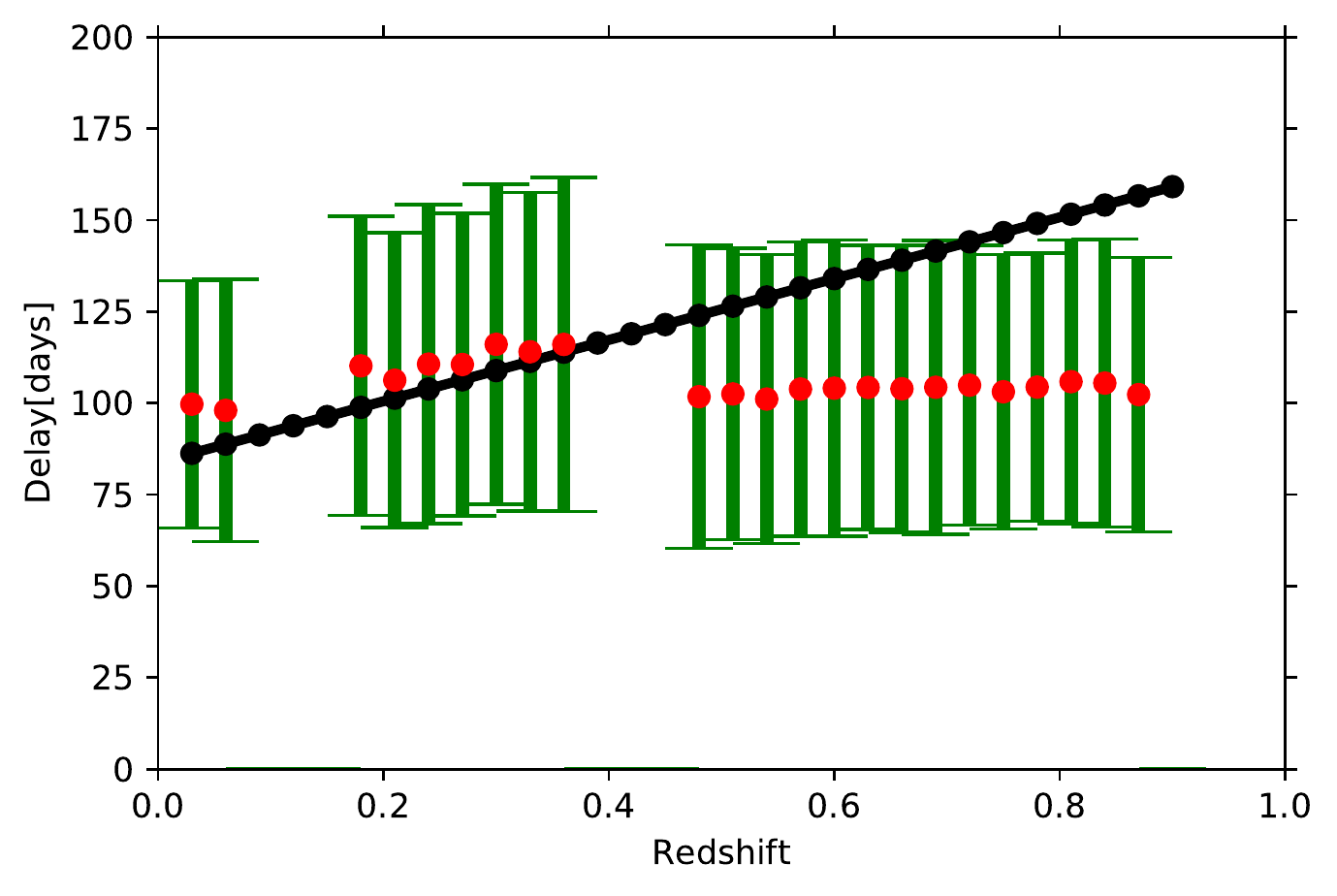}
\includegraphics[width=0.8\columnwidth]{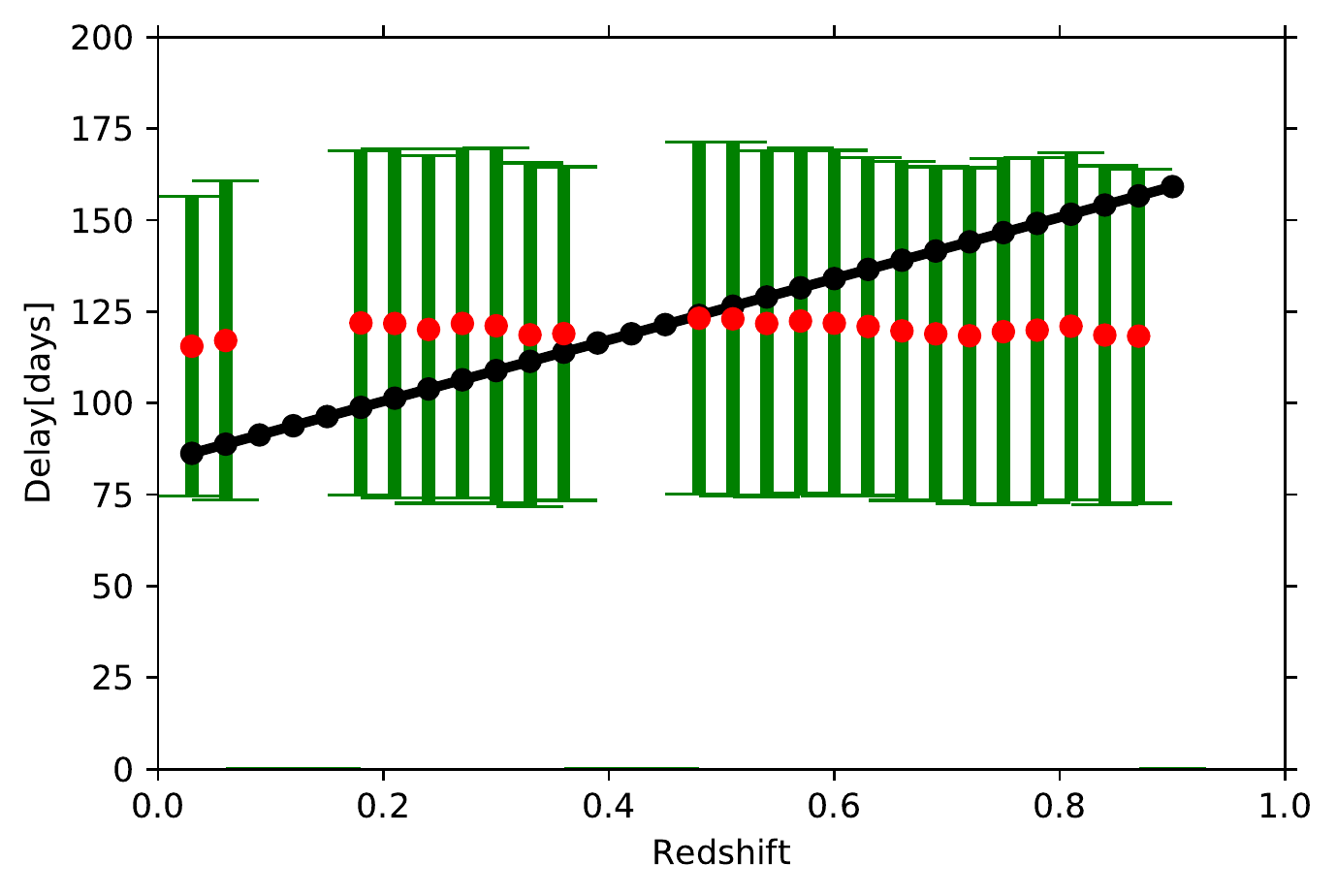}
\includegraphics[width=0.8\columnwidth]{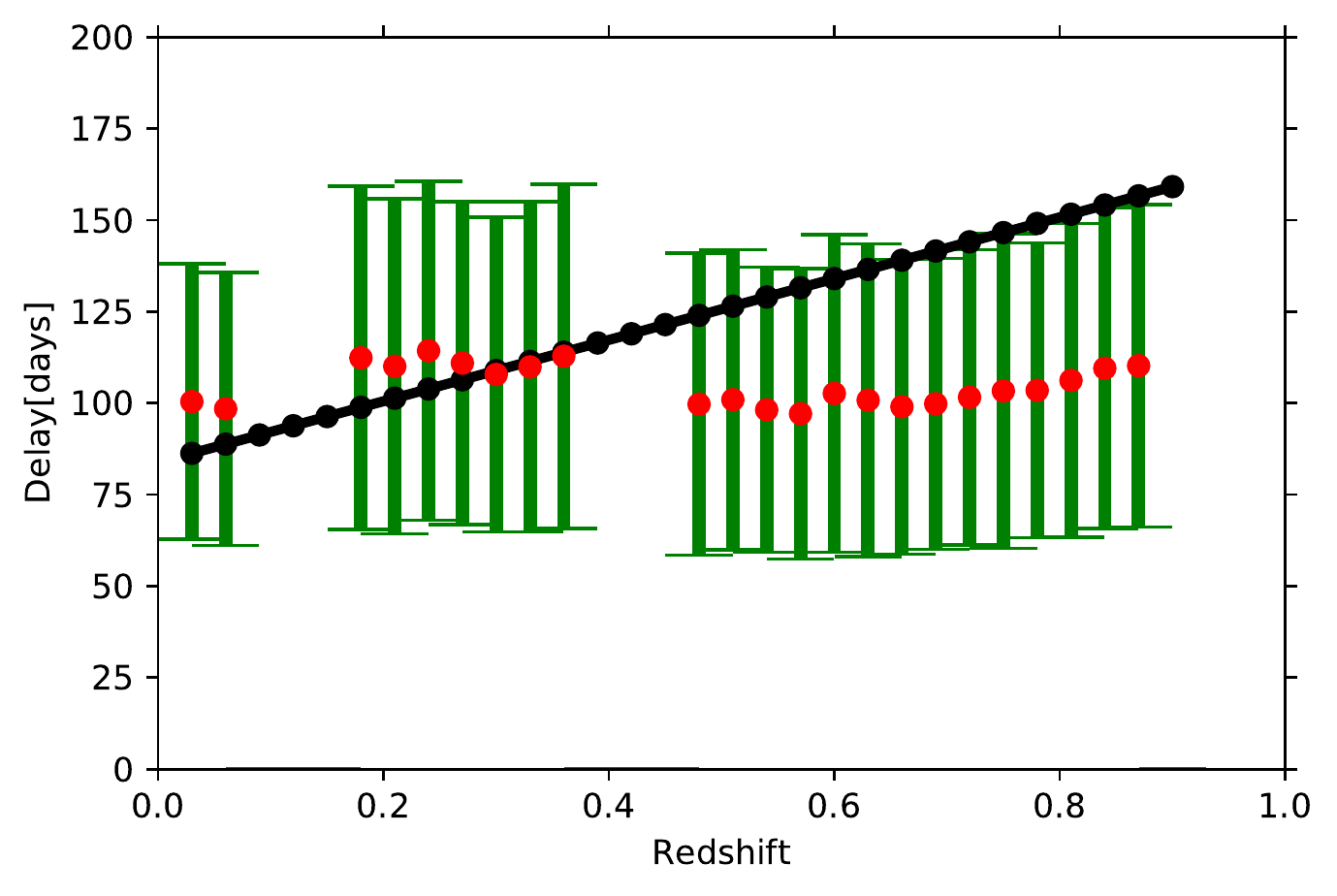}
\includegraphics[width=0.8\columnwidth]{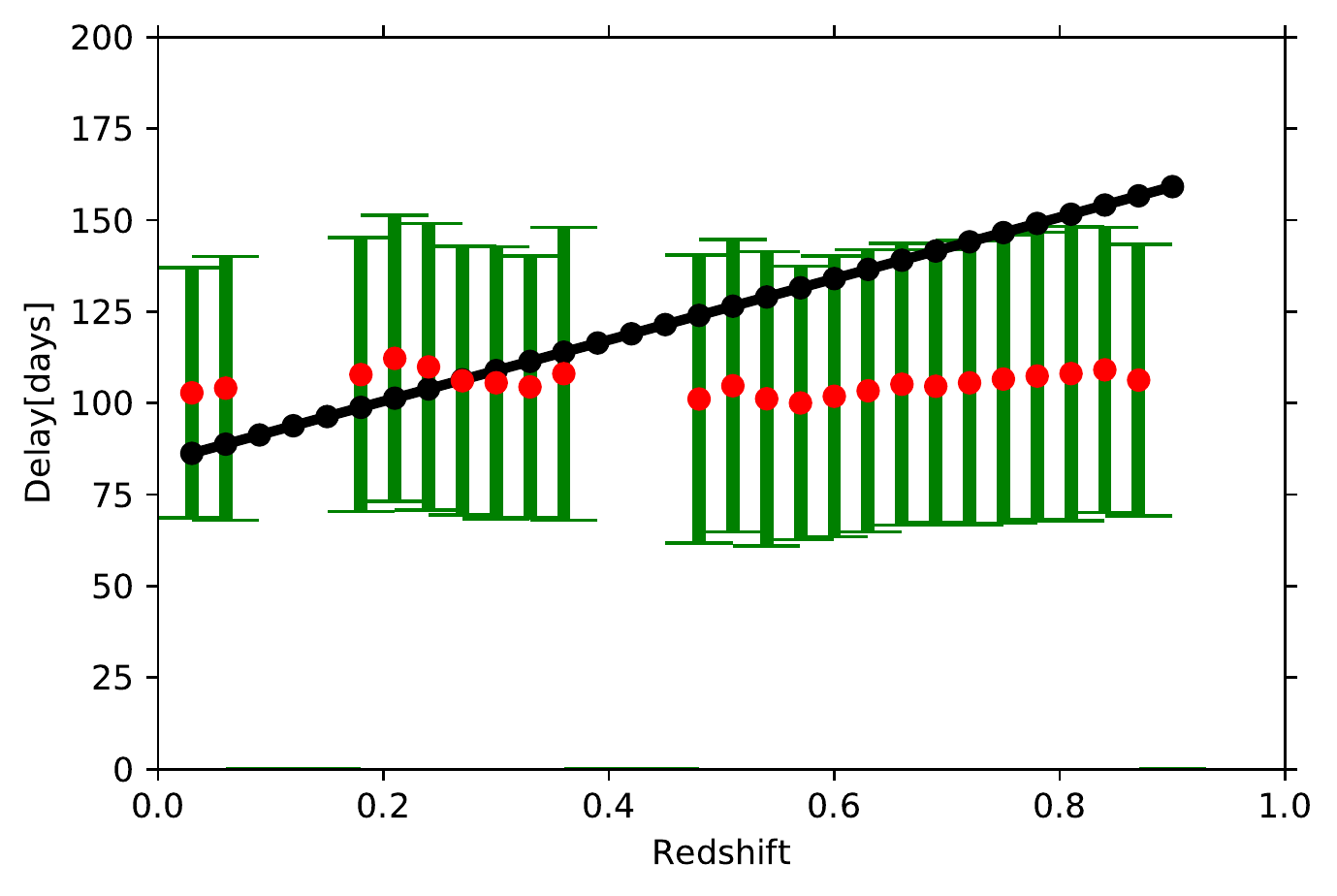}
\caption{Sensitivity of the simulations to the assumed parameter for the first year of DDF data, faint quasars $\log L_{3000} = 44.7$. Top panel: Alternative values of the line EWs. Second panel: Assuming a larger photometric error. Third panel: At shorter timescales for the frequency breaks of the power spectrum. Lowest panel: At lower variability amplitude (see Sect~\ref{sect:varia} for details).}
\label{fig:DD_various}
\end{figure}

We first tested the adopted standard values of the line EWs for the three lines (see Table~\ref{tab:parameters}). The assumed standard value for the H$\beta$ line in particular is much higher than the mean value from the SDSS given in Table~\ref{tab:table-sdss_dr14}, 150 \AA~ versus 67.36 \AA. We therefore recalculated the predictions for all the lines assuming the mean values from Table~\ref{tab:table-sdss_dr14}. We considered only the case of the faint AGN population, $\log L_{3000} = 44.7$. The result is shown in the upper panel of Figure~\ref{fig:DD_various}. Comparing the new results to the upper panel in Fig.~\ref{fig:DD1}, we 
see that the change in the EWs of the lines did not affect the results. The delay for redshifts below 0.5 would be well recovered, but higher-redshift sources are beyond the reach of the first year DDF monitoring, even for faint quasars.

We next tested the effect of the assumed photometric accuracy by replacing the rather unrealistic value of 0.001 with 0.02, hence being more conservative. This is clearly much larger than the systematic error, which is expected to be at the level of 0.005 mag \citep{LSST_Ivezic_etal_2019}. This time, we used the values of the EWs as in the previous plot, and we only changed the expected error.  The result (second panel in Figure~\ref{fig:DD_various}) is similar to the previous simulations (Figure~\ref{fig:DD1}). The delays are now determined with an error that is larger by up to 20\%, but some lo- redshift measurements are still useful. 

We also verified whether the adopted high- and low-frequency breaks are important for the simulations. We repeated the computations for the standard accuracy of 0.001, but took the high- and low-frequency breaks to be roughly ten times higher ($3.66 \times  10^{-8}$ Hz, and $3.66 \times  10^{-9}$ Hz, respectively). This caused no systematic effects and a very slight increase in the error by a few percent at most (see Fig.~\ref{fig:DD_various}). Finally, we tested the adopted level of quasar variability, but in this case, the decrease in the variability level did not affect the results of the simulations (see Figure~\ref{fig:DD_various}, lowest panel). Thus the parameterization of the variability does not seem essential for the modeling.

\begin{figure} 
\centering
\includegraphics[width=\columnwidth]{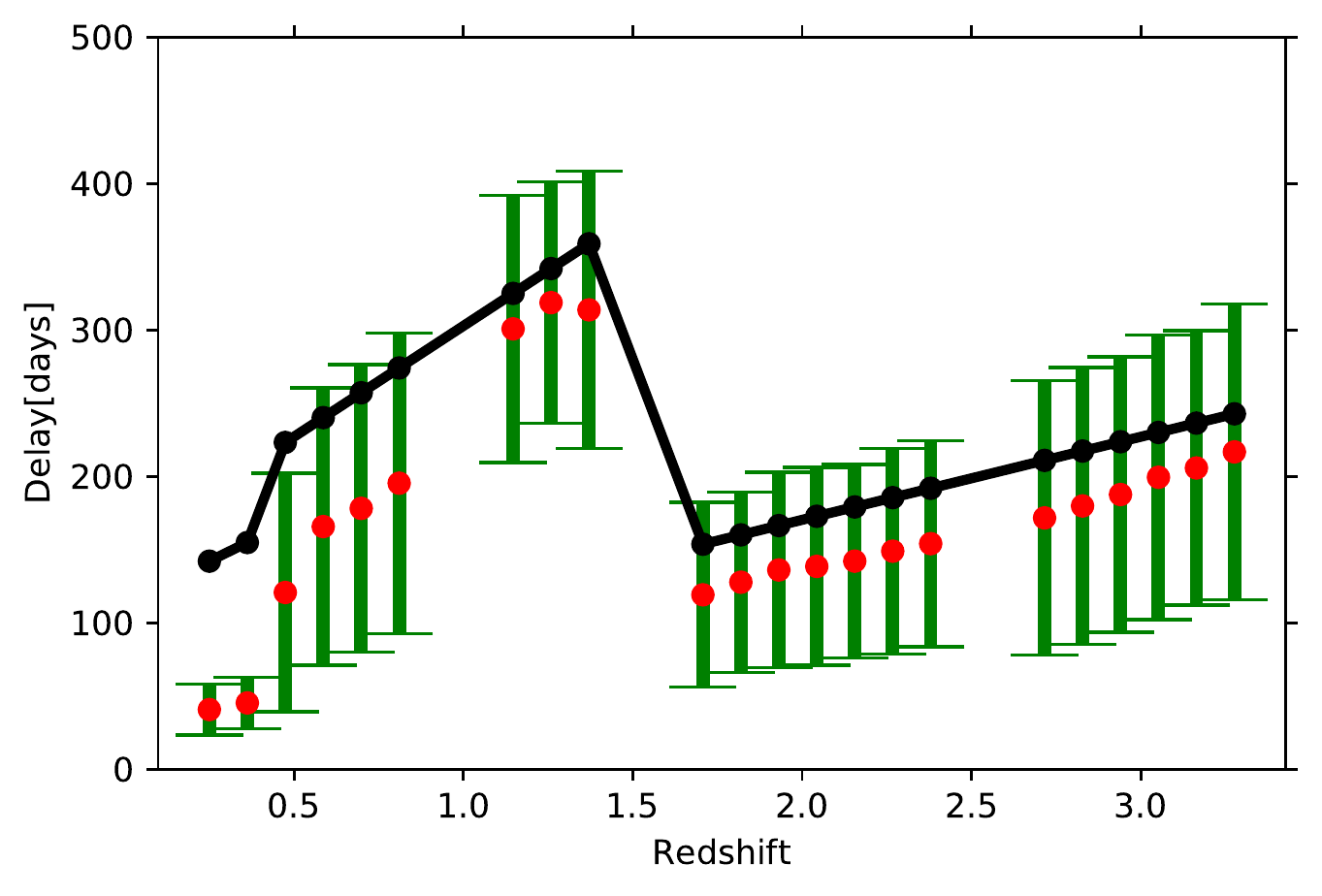}
\includegraphics[width=\columnwidth]{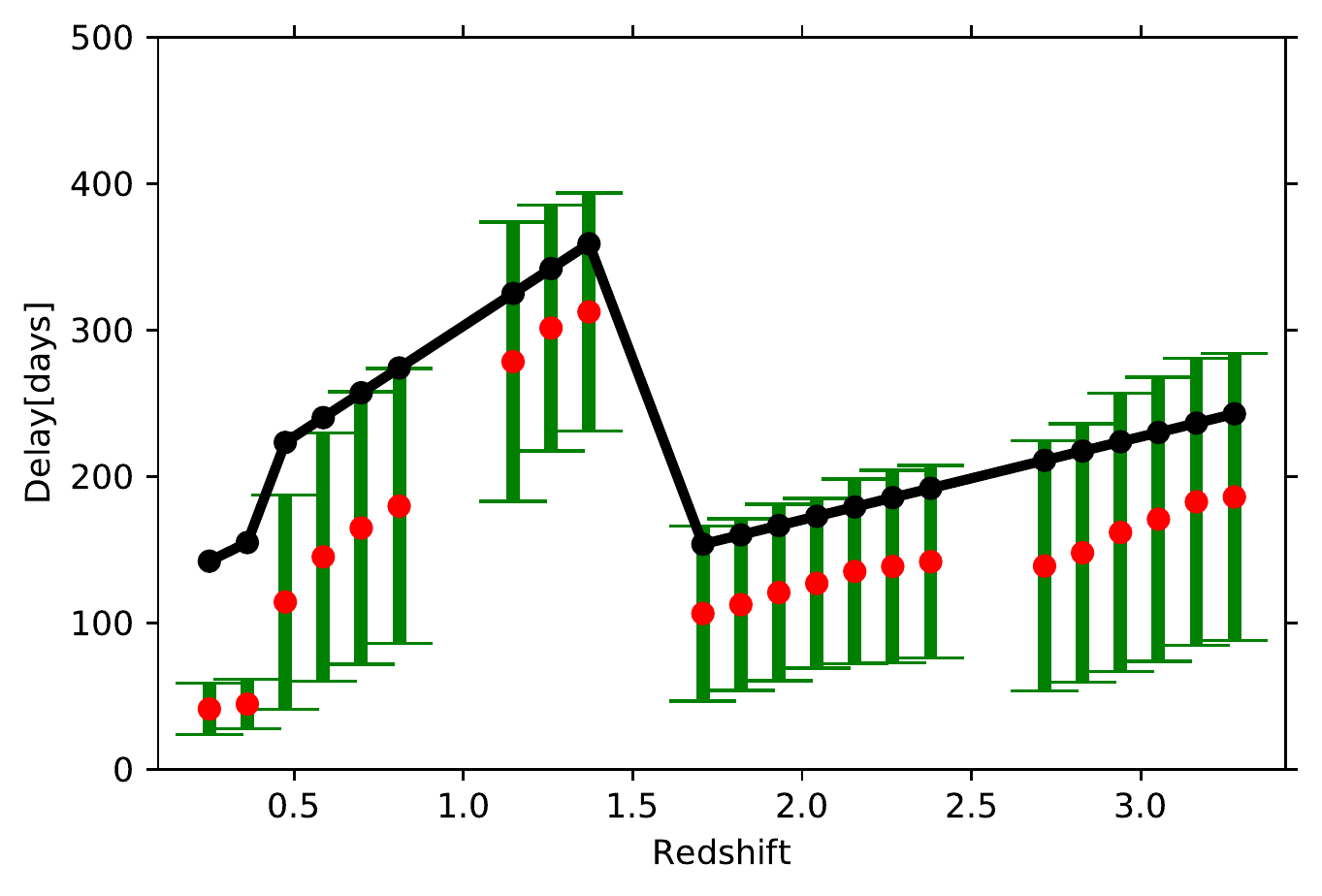}
\caption{Sensitivity of the simulations to the assumed parameter for the ten years of DDF data, bright quasars $\log L_{3000} = 45.7$, in erg s$^{-1}$, assuming a different radius-luminosity relation for each of the emission lines.  Top panel: Symmetric Gaussian response. Lower panel: Half-Gaussian asymmetric response for the BLR. 
}
\label{fig:DDF_RL}
\end{figure}

Next, we tested the effect of the assumed $R-L$ relation for the predictions. In our standard simulation setup, we used the same relation for all the lines. In order to determine whether this assumption might be problematic, we performed simulations assuming 
\begin{equation}
\log \tau (H\beta) = 1.350 + 0.415 (\log L_{3000}-44)
\label{eq:RL1}
\end{equation}
for the H$\beta$ line \citep{khadka2021},
\begin{equation}
\log \tau (MgII) = 1.67 + 0.30(\log L_{3000}-44)
\label{eq:RL2}
\end{equation}
for the Mg II line \citep{zajacek2021}, and
\begin{equation}
\log \tau (CIV) = 1.04 + 0.42 (\log L_{3000}-44)
\label{eq:RL3}
\end{equation}
for the CIV line \citep{cao2022}. The difference is most significant for the CIV line, which is suitable for quasars at higher redshifts. In this case, we therefore show the results for bright quasars from the DDF field from the entire ten years of data. This is to be compared with Figure~\ref{fig:DD10}. The new results are shown in Figure~\ref{fig:DDF_RL}. The results at low redshift changed less because they were based on H$\beta$ and Mg II, but a much shorter time delay for CIV created a problem at high redshift. It underpredicted the delay by $\sim 10$ \%. A considerable problem, however, is seen at the two lowest redshifts, below 0.5. The new formula for the H$\beta$ time delay from Equation~\ref{eq:RL1} brings values of 142 and 158 days, which are now close to half a year, and the delay determination is strongly affected by the seasonal gap. In our standard simulations, the delay was assumed to be 331 and 360 days for the corresponding two redshifts for bright quasars, so that the problem of the seasonal gap did not appear for this class of sources. 

The relations discussed above, representing the time delay as a function of the monochromatic luminosity, are not necessarily universal. There are indications that the time delays are relatively shorter for sources with a high Eddington ratio \citep[e.g.,][]{dupu2015,dupu2016}. Knowledge of the Eddington ratio, however, requires knowledge of the black hole mass and the black hole spin. In the case of spectroscopic studies, the problem can be addressed by additionally measuring the black hole mass from the line profiles or by introducing a second parameter into the relation, for example, the normalization of the Fe II pseudo-continuum, which is related to the Eddington ratio \citep{dupu2018}. This is not a good strategy for the sample, however, which is very heterogeneous \citep{khadka2022}. In the case of photometric measurements, it is not clear how the issue can be addressed in an individual object.

The next potentially important assumption was using a symmetric Gaussian to model the response of the BLR, while in the few cases, when the response function shape was derived from the data, the shape was clearly asymmetric \citep{2021ApJ...907...76H}. For the comparison, we therefore performed simulations with the response  function in the form of a half-Gaussian \citep[see, e.g.,][]{vikram2022}, but assuming the shift as implied by the formulae and the width as in a standard model, that is, 10 \% of the time delay. While we did this, we kept the line delays different for each emission line, as in the previous case. We observed that the recovery of the delay in this case is less accurate. The expected values depart by up to $\sim 20 $ \% for the adopted width of the half-Gaussian (see Figure~\ref{fig:DDF_RL}, lower panel). This systematic offset is thus a potential problem, although the recovered and expected time-delay difference is still within 1 $\sigma$ error. 

\begin{figure}
\centering
\includegraphics[width=\columnwidth]{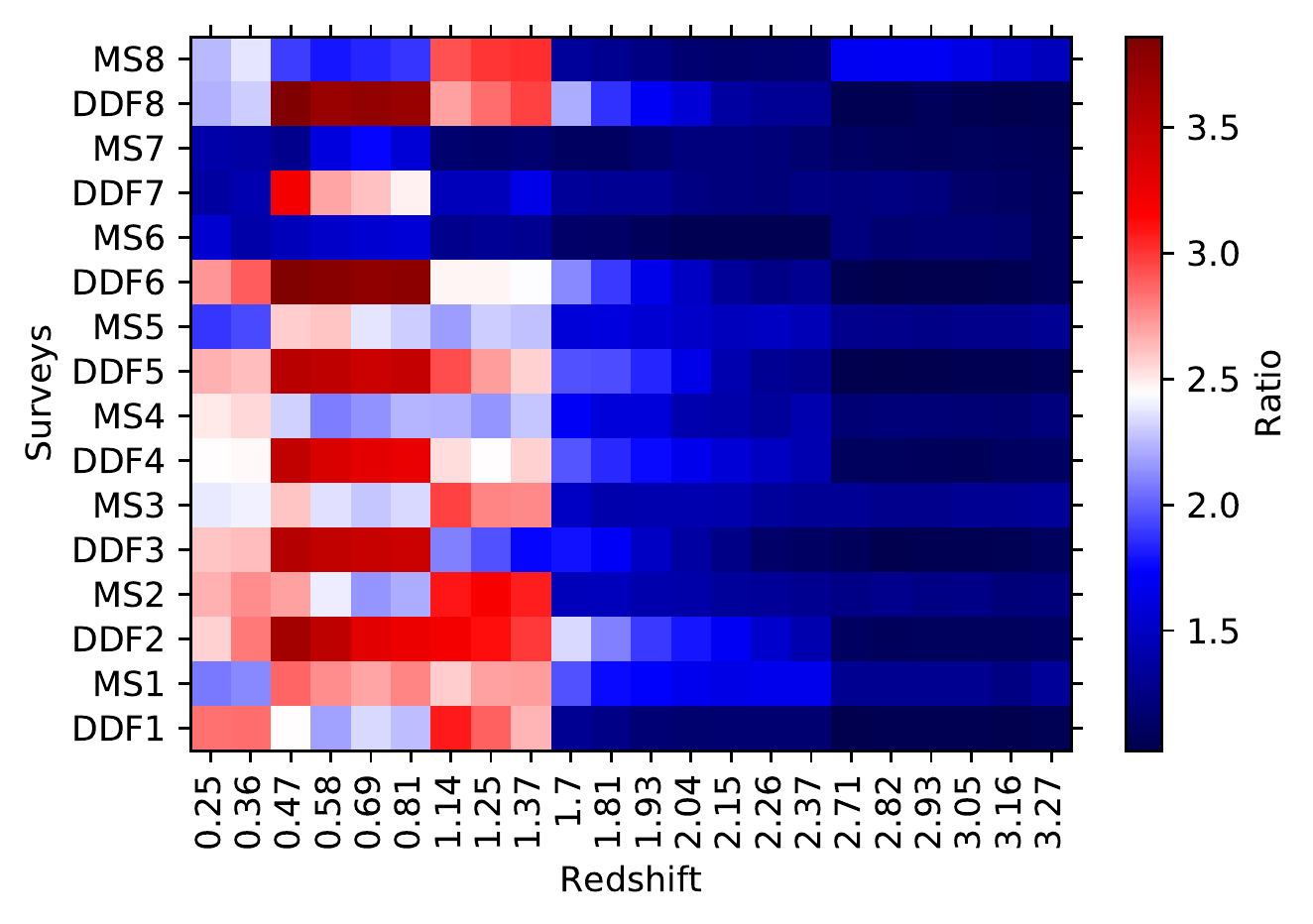}
\includegraphics[width=\columnwidth]{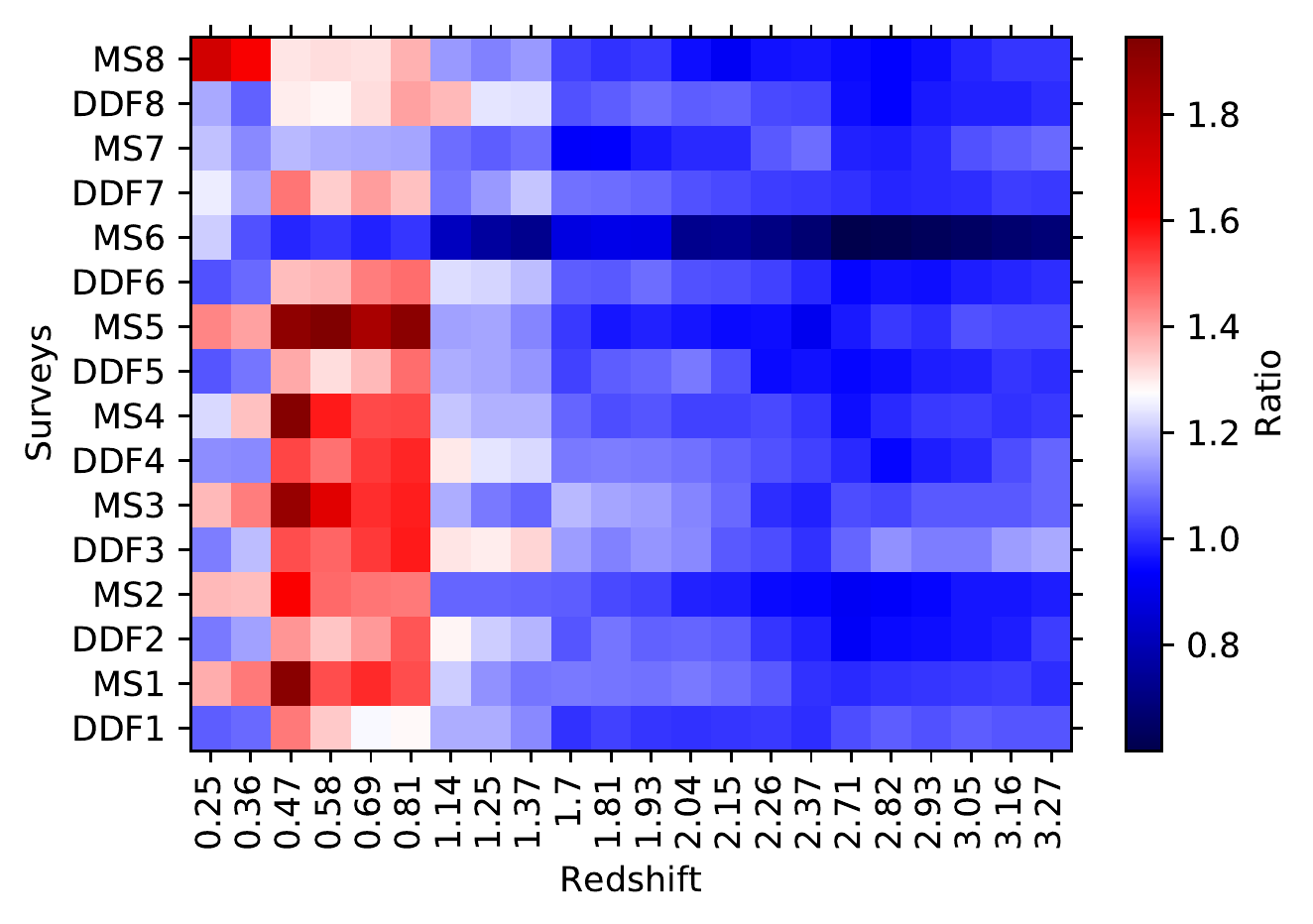}
\caption{Color-coded relative systematic error of the delay determination for faint (upper panel) and bright (lower panel) AGN in ten years of data for eight MS cadences and eight DDF cadences different from those considered before. The coding scale is different in the lower and upper panels.}
\label{fig:10yr_new_cadences}
\end{figure}

Finally, we tested other cadences than the two representative ones for MS and DDF. We used the location on the sky (0, -30) for the MS, and for the DDF, we used the ELAIS-S1 centered at (9.45, -44.025), but we tested eight more recently proposed cadences for the MS and for the DDF. We list them in Table~\ref{tab_cadence}. In order to make the presentation compact, we plot it by skipping the error bars because they do not change much. We instead show two plots that are color-coded according to the relative difference of the mean time delay, $\tau_{\rm derived}$, with the time delay adopted in the simulations, $\tau_{\rm adopted}$,
\begin{equation}
\label{eq:delta}
\delta = {\tau_{\rm derived} - \tau_{\rm adopted} \over \tau_{\rm adopted}}.
\end{equation}
This dimensionless quantity depends strongly on the source luminosity, therefore, we plot it as a function of redshift separately for faint (upper panel of Figure~\ref{fig:10yr_new_cadences}) and bright AGN (lower panel). We used all ten years of the simulated cadence. For most of the cadences, the results are overall similar to those obtained previously. The results for the bright AGN are quite satisfactory, particularly for moderate redshifts. For redshifts $ z \sim 0.7,$ the derived delays are frequently too short. As discussed already by \citet{czerny2013}, for bright quasars, we need five measurements per year if the coverage is uniform and the data is spectroscopic. For a photometric (more difficult) time-delay measurement with nonuniform sampling, we need twice that many data points, and light curves providing less than that cannot be used.
For fainter quasars, most of the cadences again underpredict the time delays, which are too short to be properly sampled in the MS mode.


In the case of DDF, the results were qualitatively similar. With the aim to compare them quantitatively, we therefore calculated two global parameters for each cadence. One parameter was the mean separation of visits after treating the observation made within one day as a single exposure. The second parameter was the redshift-averaged value of $\delta$ (see Equation~\ref{eq:delta}). In Table~\ref{tab_cadence} we list these values for the bright quasars. It is interesting to note that a simple comparison of the number of independent measurements or an actual number of visits is not reflected in the quality. For example, in the cadence S6-DDF,  the total number of visits is lower by a factor of 2 than in the remaining cadences (2542, 1672, 2637, and 2209 in $g$, $r$, $i$ $z$ bands in comparison with the mean of 5033, 2853, 5132, and 4356, respectively). However, the worst offset occurs for the cadence S3-DDF, which means that the specific distribution of observing dates is important. This offset remains, regardless of the number of observed sources. For accurate results, the offset must therefore be corrected through simulations. The large systematic error for faint sources at redshifts between 1 and 1.5 that is visible in Figure~\ref{fig:DD10} is seen in all cadences, although it is relatively smaller in S7-DDF.

\begin{table*}
\centering
\caption{\label{tab:offset}Effective mean separation in the observing dates in $r$ band and the redshift-averaged offset of the mean recovered time delay in comparison to the assumed time delay for bright quasars for ten years of data}
\begin{tabular}{cccc}
 \hline
 Cadence  &  formal name & effective  & offset  \\ 
          &              &  separation  & in delay \\
          & & [days]               &  [\%] \\
 \hline\hline
 S1-MS & \verb|baseline_v2.0_10yrs| & 13.7 & 11.7 \\
S2-MS & \verb|baseline_v2.1_10yrs| & 12.4 & 10.0\\
S3-MS & \verb|baseline_v2.2_10yrs| & 16.0 & 12.8\\
S4-MS & \verb|draft_connected_v2.99_10yrs| & 16.6 & 10.2\\
S5-MS & \verb|draft_v2.99_10yrs| & 16.2 & 11.9\\
S6-MS & \verb|light_roll_v2.99_10yrs| & 13.3 & 11.1\\
S7-MS & \verb|retro_baseline_v2.0_10yrs| & 9.4 & 7.2\\
S8-MS & \verb|roll_early_v2.99_10yrs| & 15.8 & 11.1\\
 \hline
 S1-DDF & \verb|ddf_accourd_sf0.30_lsf0.4_lsr0.5_v2.1_10yrs| & 5.8 & 8 \\
 S2-DDF & \verb|ddf_bright_slf0.35_v2.1_10yrs| & 5.0 & 12  \\ 
S3-DDF & \verb|ddf_double_slf0.35_v2.1_10yrs| & 3.2& 16 \\
 S4-DDF & \verb|ddf_old_rot_slf0.35_v2.1_10yrs| & 5.0 & 13  \\
  S5-DDF & \verb|ddf_quad_slf0.35_v2.1_10yrs| & 2.7& 10  \\
 S6-DDF & \verb|ddf_quad_subfilter_slf0.35_v2.1_10yrs| & 3.3& 10\\
 S7-DDF & \verb|ddf_season_length_slf0.20_v2.1_10yrs| & 5.9 & 10\\
 S8-DDF & \verb|ddf_season_length_slf0.35_v2.1_10yrs| & 4.7 & 11\\
 \hline
 S2-DDF-equal &  - & 1.0 & 11.1 \\
 \hline
 \hline
\end{tabular}
\label{tab_cadence}
\end{table*}

\section{Discussion}
\label{sect:discussion}

The LSST will provide up to ten million quasar detections in the main survey mode and a few thousand higher-quality AGN light curves from the DDF. It will be an enormous leap in the reverberation monitoring of AGN and for the prospects of applying the results to constrain the cosmological models. We thus performed simulations to estimate the prospects of the results from the entire ten-year survey as well as from the first two years and the first year of data collection. A large number of expected time-delay measurements will open an efficient way to study AGN properties, as well as to apply them to cosmology. We should also be aware that the large statistics of the measured time delays may reveal the dependence on the redshift, in addition to the known trends with the Eddington ratio \citep{dupu2018,panda_2022, Panda_Marziani_2022}. For example, the R-L relation may be affected by a systematic change in the average viewing angle for selected subpopulations of quasars. Currently, we do not see any such trend with the redshift \citep[e.g.,][]{prince_angle_2022,dainotti2022}, but the accuracy of this statement is low as the 95\% confidence level limit allows a change in the viewing angle from $\sim 75 $ deg (at redshift 0) to $\sim 30$ deg (at redshift 3), with a corresponding systematic change in the  luminosity distance by a factor of 1.8 (see Figure 3 in \citealt{prince_angle_2022}). 

The expected accuracy of the time-delay measurement for a single AGN is about 30 \% if the source parameters are appropriate for the measurement. During the first year, only the shortest time delays can be measured, which means that only DDF data are useful for this purpose. In addition, only faint AGN at low redshift have a time delay that is short enough to be measured. Using the SDSS AGN statistics for reference, we can expect only 15 \% of AGN to have redshifts below 0.9, and not all of them are faint. This means that overall, some 10 \% of the AGN out of $ \text{about ten}$ thousand located in the DDFs can potentially allow a measurement of the time delay. This is still a few hundred measurements and more than are available at present. It is interesting for statistical studies.  For cosmological applications, such a sample will be still too small, particularly because the dispersion in the measurements of line delays is usually high \citep[e.g.,][for the  most recent cosmological applications]{zajacek2021,cao2022}. Thus the reduction of the statistical error by a factor of 10 will not lead to high-precision cosmology yet. 

Measurements based on two years of data from the DDF will improve the situation considerably, but the critical improvement will come from the whole ten years of data. The DDF data will allow a measurement of time delays for all bright AGN and for most and even faint AGN at a redshift above 1.7. This means that at least half of the quasars in DDFs will have a determined line delay. A thousand measurements will reduce the statistical error by a factor of 30, and it will almost approach an accuracy of 1 \% in the overall distance determination. An even more spectacular improvement will come from the main survey. Only bright quasars will have delay measurements there, but this means about one million measurements. At face value, this will reduce the statistical error by a factor of 1000, which means a formal subpercent accuracy for cosmological measurements.

However, in this case, the dominant source of the error will be the systematic error. Our simulations show that this systematic error is likely to be present. Our simulations for the ten years of data in the main survey for faint AGN ($\log L_{3000} = 44.7$) always returns an average time delay that is much shorter than the assumed value  (see Figure~\ref{fig:MainSurvey}). The smallest error is for redshifts close to 3, but it is $\sim 30$\% smaller than expected. There seem to be no systematic issues with bright objects ($\log L_{3000} = 45.7$, in erg s$^{-1}$) between redshift 1.7 and 3.0. A slight deviation appears above 3, and this effect should not be present for somewhat fainter AGN because it is related to time delays that are too long compared with the survey duration. A clear discrepancy is present at the lowest redshift. However, for cosmological applications, it is necessary to cover a broad range of redshifts, including the lowest redshifts. A further study of the systematics present in this case below 1.7 is therefore necessary. The solution may be to use extensive modeling and the appropriate corrections, but they will most likely depend on the specific source luminosity and the actual cadence in the field. Combining the results from the DDF and the main survey will also improve the situation because for DDFs, the results at lower redshifts are more accurate. As mentioned above, the additional intrinsic systematic trends in AGN may also be detected. Separate studies of subclasses of AGN are therefore required. In order to achieve high accuracy of the results, the intrinsic time delays between the continuum bands should also be included in the actual data analysis. The potential errors related to this issue were not estimated in the current study.

The cadence adopted in the simulations is one of the main sources for the offset. We checked that by repeating the computations for the faint quasars, two years of data, assuming the same number of observations as implied by the DDF cadence used in Figure~\ref{fig:DD2}, but assuming a seasonal gap of 180 days between years 1 and 2, and roughly equally spaced observations during the remaining six-month periods. In this case, the agreement between the assumed and recovered time delays is much better, although the dispersion in the redshift recovery is still roughly the same. However, this cadence is not likely to be accepted during the DDF monitoring. Otherwise, we do not see high sensitivity of the line delay recovery on the tested cadences that were considered here. The possibility of measuring the time delay mostly depends on the source luminosity and redshift, and not as much on the actual cadence. 

This is clearly different from the measurement of the continuum time delay, where dense sampling is essential, as argued already by \citet{brandt2018}. For this reason, \citet{LSST_Kovacevic2022} argued that useful measurements can be mad only for DDF, and some of the cadences were favored. Simulations performed by \citet{pozo2022} also request a two- to five-day cadence for continuum delays with the LSST, and such cadence is roughly expected in DDF fields although not forming such a regular pattern as assumed in that paper.
 
Our simulations are based on the creation of the light curves using the \citet{TK1995} algorithm. This is not the only method available, although it is fairly general due to a number of parameters entering into the parameterization of the power density spectrum. Other methods of creating artificial light curves are also used, mostly working directly in the time domain, such as the damped random walk \citep{kelly2009}, the damped harmonic oscillator \citep{DHO2022}, or more complex higher-order CARMA processes \citep{kelly2014}. These methods were used by \citet{2022MNRAS.512.5580S} to simulate quasar light curves in the DDF and aimed at the precise reconstruction of the light curves from the available photometric data. They used the advanced method of stochastic recurrent neural networks and concluded that the recovery precision is most affected by the seasonal gaps. In our simulations, we neglected the narrow components of the emission lines, but for quasars, the narrow components are usually relatively weak. 

More importantly, we assumed in the present simulations that we know the luminosity and the redshift of each source. The redshift is important both for the position of the emission lines and also for the luminosity and the estimate of the expected time delay.  However, most of the quasars, particularly the fainter sources, will be discovered in the course of the survey, which poses a problem for their identification and the photometric measurement of the redshift. The automatic AGN classifier (see \citealt{russeil2022}) is already working in the case of the Zwicky Transient Facility, and a similar classifier is under development for the LSST. The accuracy of the photometric redshifts for AGN is not well established, therefore we did not attempt to model this effect, but it will be likely a considerable source of error in the MS, where spectroscopic follow-up of a large fraction of sources is not realistic. If the photometric errors are small, it would be enough to broaden the redshift ranges, which should be avoided because the band does not contain a full line, which would simply reduce the number of suitable objects.




\section{Conclusions}
\label{sect:conclusions}

We showed that the recovery of the emission line time delay from the photometric measurements available from the LSST is possible for a significant fraction of the quasars. The expected time delays depend on the source luminosity, but quasars typically range from $\sim 100$ days to over three years, so that the specific cadence requirements are not as critically important as for continuum time delays. We list our results below.

\begin{itemize}
\item For quasars brighter than $\log L_{3000} = 45.7,$  the cadence available in the main survey is in general good enough to allow measurements of the line time delay with respect to the continuum; individual measurement errors are large, about 40 \%, but in most of the cadences, the systematic offset is 10 - 15 \%. Combining  measurements for many quasars will therefore allow us to  statistically study trends such as the  radius-luminosity relation.\item For quasars fainter than $\log L_{3000} = 44.7$,  the main survey is not recommended, and for the intermediate-luminosity quasars, the redshift limit of practical use will have to be set through simulations.
\item The line time delays in quasars fainter than $\log L_{3000} = 44.7$  can be successfully measured from the DDF. In this case, even the first two years of data are enough, and the longer data set does not improve the delay measurement at lower redshifts unless the data are sampled in shorter periods or are detrended.
\item Bright quasars can be also studied with dense sampling when they are located in DDF fields, but this only slightly decreases the individual error (down to below $\sim 30$ \%).
\item Each of the considered cadences leads to some systematic offset between the delay assumed in the setup and the recovered time delay. This offset (about 10 \%)  will remain even if numerous quasars are measured. High-quality results therefore require correcting for this offset by numerical simulations. This offset will depend on quasar properties as well as cadence, photometric errors,  and the time-delay measurement method. 
\item Some of the considered cadences are better than others, but for the line-delay measurements, the cadence is apparently not a critical issue.
\end{itemize}

Therefore, with proper selection of the source luminosities and corresponding redshift ranges, we expect reliable measurements of the time delays for a significant fraction of the quasars observed in the main survey mode when the project is completed. The exact fraction could be estimated when the observational cadence is set. Results for intrinsically fainter sources located in DDF fields and with low redshifts can be obtained even from the data collected during the first year of the project.

\section*{Acknowledgements}
This project has received funding from the European Research Council (ERC) under the European Union’s Horizon 2020 research and innovation program (grant agreement No. [951549]).
The project was partially supported by the Polish Funding Agency National Science Centre, project 2017/26/\-A/ST9/\-00756 (MAESTRO  9), project 2018/31/B/ST9/00334 (OPUS 16), and OPUS-LAP/GAČR-LA bilateral project (2021/43/I/ST9/01352/OPUS 22 and GF23-04053L). Partial support came from MNiSW grant DIR/WK/2018/12. This project has received funding from the European Research Council (ERC) under the European Union’s Horizon 2020 research and innovation program (grant agreement No. [951549]). SP acknowledges the Conselho Nacional de Desenvolvimento Científico e Tecnológico (CNPq) Fellowships 164753/2020-6 and 300936/2023-0. MZ, RP, and BC acknowledge the Czech-Polish mobility program (M\v{S}MT 8J20PL037 and PPN/BCZ/2019/1/00069). SK acknowledges the National Science Centre grant (OPUS16, 2018/31/B/ST9/00334). 
A.B.K., D.I., and L.\v{C}.P. acknowledge funding provided by the University of Belgrade—Faculty of Mathematics (contract No. 451-03-47/2023-01/200104) and Astronomical Observatory Belgrade (contract No. 451-03-47/2023-01/200002) through the grants by the Ministry of Science, Technological Development and Innovation of the Republic of Serbia. D.I. acknowledges the support of the Alexander von Humboldt Foundation. M.L. M.-A. acknowledges financial support from Millenium Nucleus NCN$19\_058$ (TITANs).
FPN gratefully acknowledges the generous and invaluable support of the Klaus Tschira Foundation. MZ acknowledges the financial support from the GA\v{C}R-LA grant no. GF22-04053L. 

\bibliographystyle{aa}
\bibliography{main}

\clearpage
\onecolumn

\appendix

\section{Error of the delay measurement in simulations}

\subsection{Exemplary histograms of the time delay}
\label{sect:histo_100_1000}

As outlined in Section~\ref{sect:stat}, we used 100 random realizations of the process to determine the time-delay accuracy. The mean time delay is determined as the average of the 100 delays, and the error as the dispersion. This is a very simple procedure, but a more advanced procedure seems to be problematic because the time delay of the BLR lines measured from the photometry is difficult. 

\begin{figure}[!htb]  
\includegraphics[width=\columnwidth]{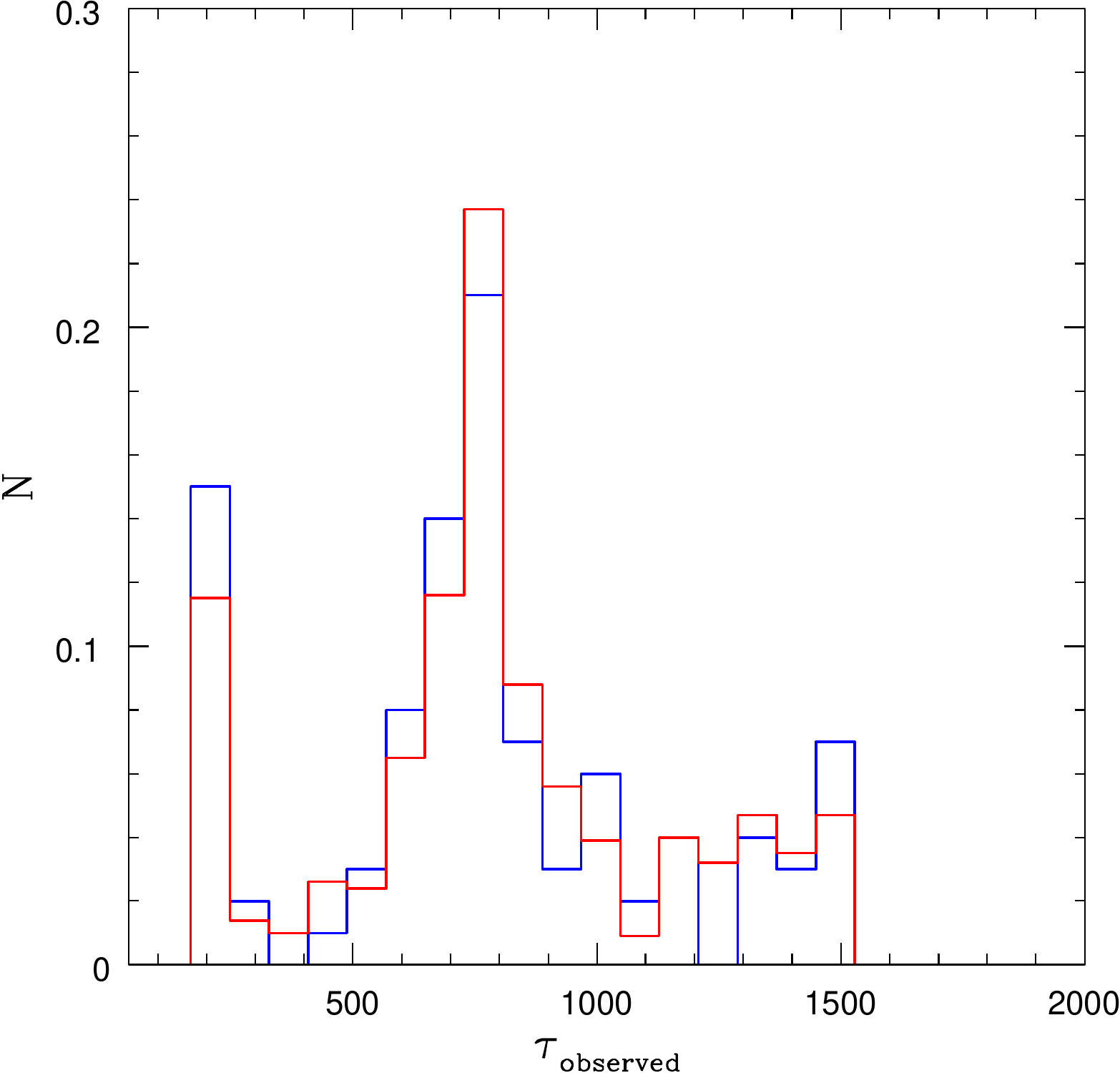}
\caption{Example of the histogram of the 100 random realizations for the time delay (blue histogram), and 1000 realizations of the same process (red histogram). The assumed redshift is 1.93, and the assumed luminosity is $\log L_{3000} = 45.7$, $L_{3000}$ in erg s$^{-1}$,. The expected time delay is 776.27 days.}
\label{fig:histoplot}
\end{figure}

We illustrate the distribution of the 100 time delays in Figure~\ref{fig:histoplot}. The computations were made for the standard model, sain Survey cadence, redshift $z = 1.93$, and a quasar luminosity of $\log L_{3000}= 45.7$, $L_{3000}$  in erg s$^{-1}$. The delays do not have a Gaussian distribution. Instead, a few peaks are visible.  The mean of the derived time delays for the adopted parameters is 778.57 days, while the maximum in the histogram is at 768.12 days in the observed frame. The dispersion calculated from the delays gives 361.99 days, which would imply the results for the time delay with 1 $\sigma$
error between 416.58 days and 1140.56 days. When we determine the 1 sigma error from the actual complex histogram, the corresponding numbers are 241.87 days and 1131.30 days. The error is therefore somewhat larger due to the complex shape. However, the representative values can be derived from our simplified approach. 

We repeated the same simulation, but using 1000 realizations instead of 100. The results are also plotted in Figure~\ref{fig:histoplot}. In this case, we obtain the mean delay of 799.58 days, with a dispersion of 340.19 days. The expected dispersion is large. The 1 sigma result is between 427.96 days and 1174.56 days when it is determined from the histogram. A larger number of statistical realizations does not change the histogram essentially. A secondary peak again appears at the shortest delays as well as at the longest delays, but the central peak is close to the expected value. 

The histogram has additional peaks  at the highest and the longest delays because the search is performed in a limited range of likely delays. If no range in the search is imposed, the best delay value frequently peaks at zero delays or at the longest wavelengths. The current version of the code for each light-curve set selects only the single best value for the delay. A more advanced version should store three (or more) values and then select the most likely value a posteriori. This certainly should be done when analysing actual data.

\subsection{Multiple characters of the best-fit solutions}

\begin{figure}
\centering
\includegraphics[width=\columnwidth,trim={1cm 2cm 1cm 5cm},clip]{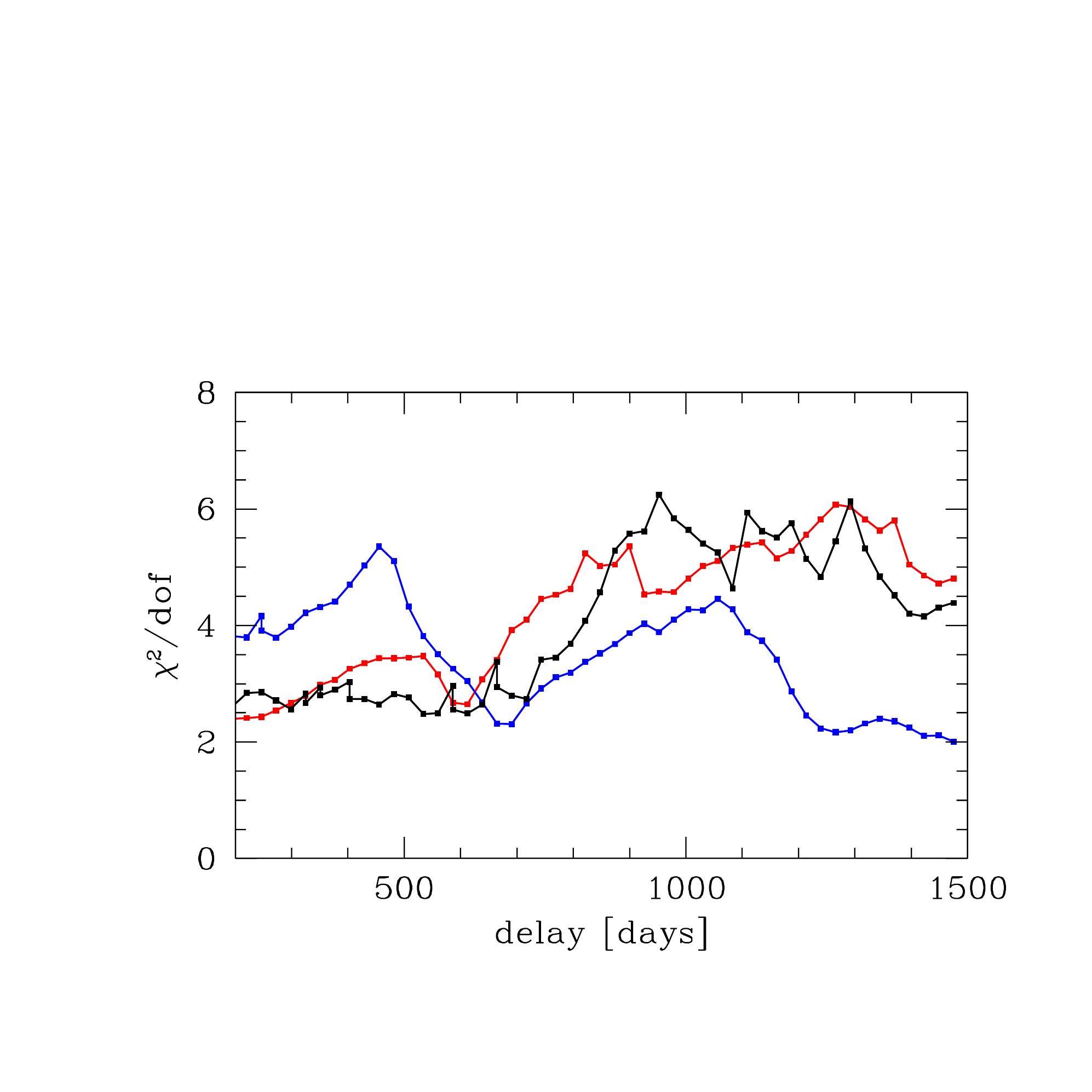}
\caption{ $\chi^2$ distribution as the function of the assumed time delay for three exemplary light curves showing the best-fit time delay of 533 days (black line), 1475 days (maximum allowed; blue line), and 194 days (minimum allowed, red line). They all show a minimum around $\sim 700$ days, but this is only a local minimum. The assumed time delay is 776.27 days.}
\label{fig:chi2_dist}
\end{figure}

As mentioned in Section~\ref{sect:histo_100_1000}, the best-fit solutions peak at the minimum and the maximum allowed value of the delay during the period search. This is related to the complex character of the function that implies the best fit. In the case of the $\chi^2$ method, the best fit corresponds to the minimum value, and the function frequently has a number of minima. We took three of the special solutions shown in Figure~\ref{fig:histoplot} selected in such a way that one has the minimum roughly where expected, while the other two have the minimum at the shortest and the longest delays, respectively. We plot the $\chi^2$ distribution for these three cases in Figure~\ref{fig:chi2_dist}.

Assuming a narrower search range for the delay can automatically reduce the error, but in the case of the actual single datum for a quasar, we will have only a crude estimate of the expected time delay from its redshift and magnitude. The luminosity estimate may additionally be biased by the intrinsic extinction in the source. We thus tested whether confirming the correlation between the two cross-correlated curves helps to eliminate sources that likely lead to the deepest minimum at the wrong location.

\subsection{Curve quality and optional curve preselection}

\begin{figure}  
\centering
\includegraphics[width=\columnwidth,trim={1cm 2cm 1cm 5cm},clip]{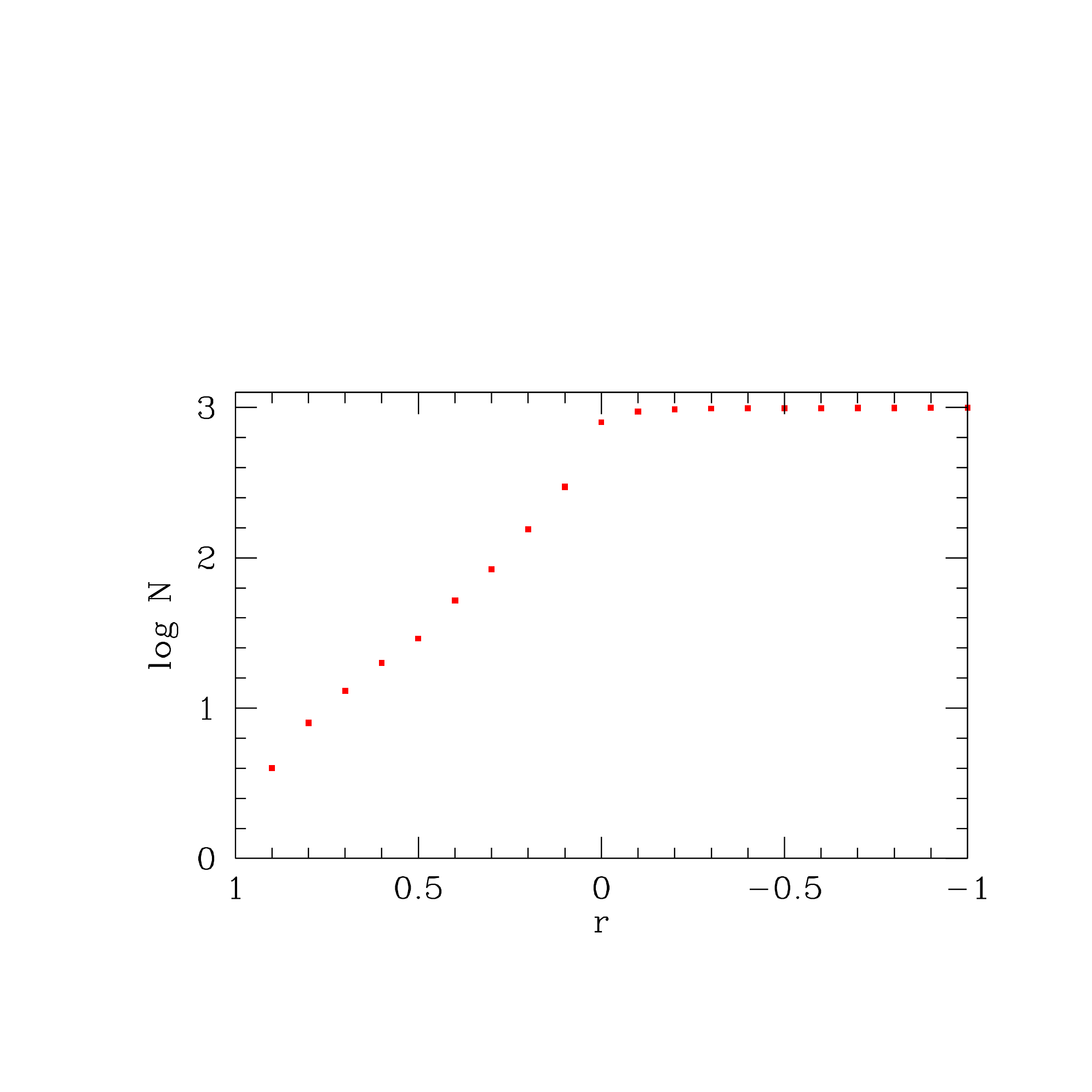}
\caption{Number of curve pairs left out of 1000 simulated pairs as a function of the adopted cutoff for the correlation coefficient $r$. Only 84 pairs out of 1000 have $r > 0.3$. }
\label{fig:rr_ile}
\end{figure}

We used the 1000 simulated realizations shown in Figure~\ref{fig:histoplot} in order to assess the quality of the quasar curves in the MS.  After optimized band subtraction as explained
in Eq.~\ref{eq:subtraction}, we calculated  for each of the created pairs of curves the correlation coefficient $r$ with the aim to determine whether a pair preselection can improve the accuracy of the delay determination for most quasars without the loss of many sources. In Figure~\ref{fig:rr_ile} we show the number of pairs that is left as a function of the choice of minimum acceptable value of $r$.   Less than 10\% of quasars (for the adopted luminosity, redshift, and MS observational pattern) may in principle guarantee reliable measurements. Overall, the curve quality is low, but our simulations most likely represent the actual data quality well. Still, having a million objects or more, we can have encouraging statistics if the results are not biased.

\begin{figure}
\centering
\includegraphics[width=\columnwidth,trim={1cm 2cm 1cm 5cm},clip]{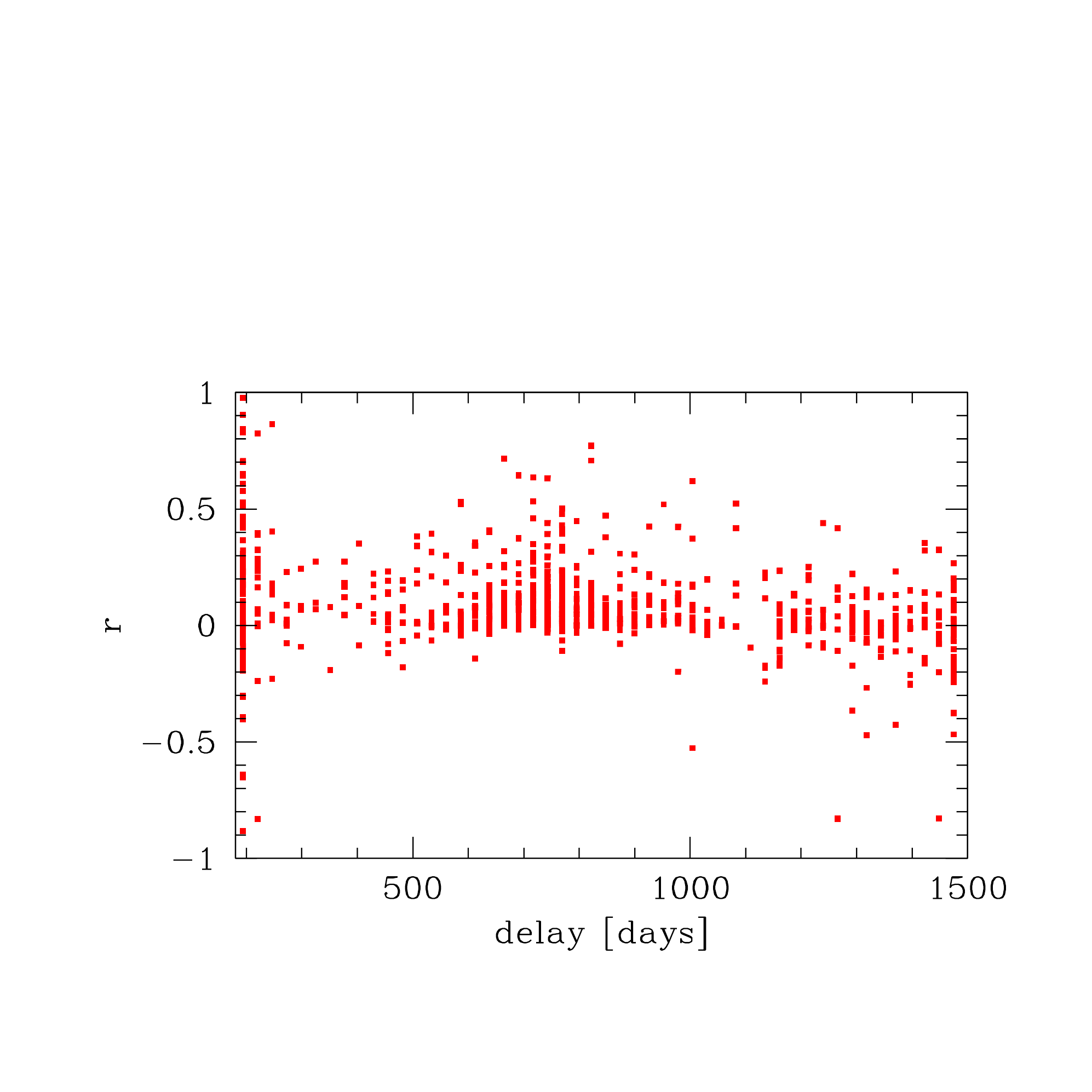}
\caption{Values of the correlation coefficients $r$ for 1000 pairs as functions of the measured time delay. High values of $r$ are not only close to the expected time delay, but lie mostly around the lower limit adopted for time-delay measurements.}
\label{fig:correlation}
\end{figure}

However, selecting realizations (sources) only according to their correlation coefficient does not lead to the successful removal of the wings seen in Figure~\ref{fig:histoplot}. In Figure~\ref{fig:correlation} we show all the values of the 1000 correlation coefficients as a function of the time delay predicted for the given pair. Many of the highest values populate the region in which the time delays are too short. This is expected. Part of the effect is due to the imperfect subtraction of the bands with and without an emission line, which is well modeled in our software. This will be surely present in the actual LSST data. The other aspect may be related to the red-noise character of the quasar light curves. We observed strong local minima for unexpectedly short time delays in our observational spectroscopic studies of the quasar time delay with the Southern African Large Telescope, where the cadence was similar to what is expected from the LSST \citep{czerny2019,zajacek2020,zajacek2021,prince_CTS_2022,Prince_etal_2023}. 

\begin{figure}
\centering
\includegraphics[width=\columnwidth,trim={1cm 2cm 1cm 5cm},clip]{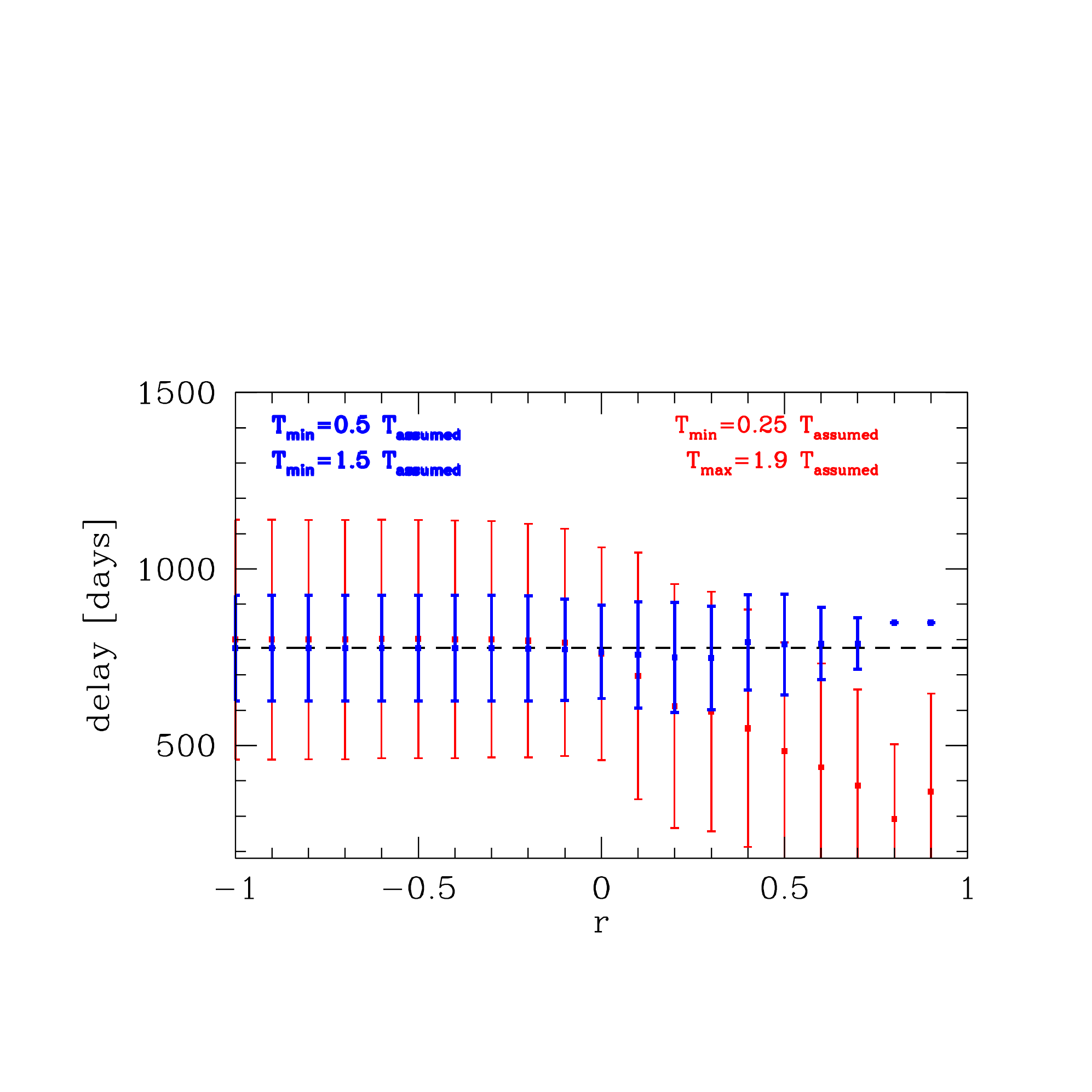}
\caption{Time delay from the subsample of 1000 simulations selected according to the minimum value of $r$ (simulations with the maximum value of the correlation coefficient smaller than r were removed from the sample) and the adopted limits for time delay. Narrower limits clearly give smaller errors, which is reflected in the dispersion. }
\label{fig:delay_rr}
\end{figure}

Therefore, we performed a combined study of the selection based both on the value of $r$ and the lower limit for the time delay. In the minima of the histogram~\ref{fig:histoplot}, we selected the minimum and the maximum time delays as 0.5 $T_{assumed}$ and 1.5 $T_{assumed}$, respectively, and we repeated the previous study. The error bars are clearly smaller for any constraints on the minimum value of $r$, and they are particularly reduced for $r > 0.7$, down to 75 days, but only four objects satisfy the two criteria of 1000 simulated pairs. For a conservative choice of $r > 0.3$, the mean value and the dispersion is $748 \pm  146$ days, and we have 50 such pairs. The overall trend with the change in the limiting value of $r$ is shown in Figure~\ref{fig:delay_rr}. We also considered a recalculation of the time delays assuming the same lower and upper limit for a delay search in the code itself, instead of selecting suitable pairs a posteriori. However, running a new set of 1000 simulations with the imposed narrower search for the delay did not improve the situation much. For a conservative case of $r > 0.3$, the mean value of the dispersion is $ 681 \pm 170$ days, and 73 such pairs satisfy this constraint. By selecting 50 out of 1000 cases, we thus reduce the dispersion by a factor of 2.3 (from 340 down to 146 days), but if we were to create 50 bins from the original 1000 objects, then the dispersion would decrease by a factor $\sqrt{20}$. Overall, it therefore seems a better strategy to use all the data and finally bin them. At some level, the systematic effects might become essential, as indicated by the differences between the {\it \textup{mean}} values from the simulations and the assumed values in the simulations, as presented in the figures in Section~\ref{sect:results}. 

\subsection{Grouping quasars in the actual data}

The errors in a single measurement shown in Figures~\ref{fig:MainSurvey} - \ref{fig:DDF_RL}  are large, but a preselection of the sources (except for selecting only a suitable range of redshifts and source luminosity, which does not give an offset between the assumed and the mean recovered value of the time delay) does not seem to give an efficient way to reduce the error. However, if the aim is not an individual source, but the construction of radius-luminosity relation or luminosity distance measurements, the efficient way is to group the measurements. 

In the case of the MS survey, we will have at our disposal about one million objects with measured emission line delays. These measurements can be grouped into relatively narrow redshift bins as well as into luminosity bins and then averaged. This should provide an efficient way to reduce the error when establishing a mean trend.

In the case of DDF fields, there is an additional interesting option for fainter sources at low redshift. Since we showed in Figure~\ref{fig:DD1} that for sources like this, one year of data is enough to perform the measurements, the delay can be measured in ten consecutive years independently after the survey is completed, and can then be combined into a single time delay, with additional quality control from the measurement dispersion. This quality control will be essential because we know well that some nearby sources can show large variations in the measured H$\beta$ line delay between the consecutive years, with more complex scaling with luminosity than in statistical samples of different sources \citep[e.g.,][for NGC 5548]{peterson2002,lu2016,2021ApJ...907...76H}. 

\subsection{Time-delay measurements using Javelin}

We also tested the use of the Javelin code to determine the time delay \citep{2011ApJ...735...80Z,2013ApJ...765..106Z}. We focused on the simulations for the MS for bright quasars. When the expected time delay was allowed to be between 0 and 3000 days for all the redshifts, the measured time delay was uncorrelated with the assumed time delay, and no clear trend with the redshift was seen, in contrast to expectations (the assumed rest-frame time delay was redshift independent, but the curves are analyzed in the observed frame). Next, we applied the same restrictions for the time-delay search as in the $\chi^2$ method. The minimum value was adopted as 0.25 of the assumed delay in the observed frame, and the maximum value was determined as the smaller value of the two: (a) half of the monitoring period, and (b) 1.9 times the expected time delay. For each redshift we performed 10 statistically independent simulations, i.e., we used ten different (but statistically equivalent) curves created with the \citet{TK1995} algorithm.

We compared in detail the results for the redshift z = 0.25 and the redshift z = 3.27. For the smaller redshift, we obtained  the mean delay from 10 realizations: 285.9 days, the mean error reported by Javelin is 82.1 days (calculated by averaging all reported positive and negative time delays), and the mean dispersion between the best fit time delays is 149.1 days. The dispersion between different realizations of statistically identical curves was larger than the error typically reported by Javelin. For the redshift z = 3.27, the values are the following:
mean delay = 1211.6 days,  mean error = 160.5 days, and mean dispersion = 179.8 days. In this case, the mean error and the mean dispersion are similar. The mean time delays for $z = 0.25$ in both Javelin and the $\chi^2$ methods are shorter than assumed in the modeling, but within the error (285.9 days and 218.1 days, respectively, vs. 331.0 days assumed in the simulations). For the high redshift $z = 3.27$, the recovered mean time delays (Javelin: 1211.6 days, and $\chi^2$: 1041.8 vs. the assumed value of 1132.4 days) were also consistent with the assumed values. Javelin, except for some underestimation of the error, is also a viable method, but much more computationally intensive.

\section{Details for the cadence simulations}

A brief description for each of the selected cadence simulations used in this paper is summarized in Table \ref{tab:lsst-description}. For more details, we refer to the official {\sc rubin\_sim} webpage\footnote{\url{https://rubin-sim.lsst.io/}}.

\begin{landscape}
\begin{longtable}[c]{lllll}
\caption{Details of the cadence simulations from the rubin\_sim}
\label{tab:lsst-description}\\
\textbf{ID} &
  \textbf{run} &
  \textbf{brief} &
  \textbf{family} &
  \textbf{description} \\ \hline\\
\endfirsthead
\caption{continued}\\
\endhead
\hline\\
\endfoot
\endlastfoot
S1-MS &
  baseline\_v2.0\_10yrs &
  v2.0 baseline &
  baseline &
  \begin{tabular}[c]{@{}l@{}}The v2 baseline simulations. V2.0 and V2.1 feature \\ a survey footprint with expanded dust-free area and \\ WFD-level visits in the Galactic Bulge and Magellanic \\ Clouds. Coverage of the Northern Ecliptic Spur, South \\ Celestial Pole, and remainder of the Galactic Plane is \\ maintained, at lower levels. Filter balance is modified \\ in different areas of the sky. A 2-region rolling cadence \\ is used in the dust-free WFD sky, to improve cadence. \\ The v2.1 simulation adds coverage in the Virgo cluster \\ and acquisition of good seeing images in r \& i bands. \\ The retro simulations are intended to provide a bridge \\ from v1.X to v2.X, introducing modifications to the \\ footprint and scheduler code separately and in stages.\end{tabular} \\\hline\\
S2-MS &
  baseline\_v2.1\_10yrs &
  \begin{tabular}[c]{@{}l@{}}v2.1 baseline \\ (+Virgo, good seeing)\end{tabular} &
  -''- &
  -''- \\\hline\\
S3-MS &
  baseline\_v2.2\_10yrs &
  \begin{tabular}[c]{@{}l@{}}v2.2 baseline \\ (prescheduled DDFs)\end{tabular} &
  -''- &
  -''- \\\hline\\
S4-MS &
  draft\_connected\_v2.99\_10yrs &
  \begin{tabular}[c]{@{}l@{}}DDF at 7.5\%, twilight \\ NEO microsurvey, \\ priority GP footprint.\end{tabular} &
  draft v3 &
  \begin{tabular}[c]{@{}l@{}}Draft versions of simulations incorporating the \\ SCOC Phase 2 recommendations for survey strategy.\end{tabular} \\\hline\\
S5-MS &
  draft\_v2.99\_10yrs &
  \begin{tabular}[c]{@{}l@{}}Less connected, agg\_level 1.5, \\ galactic plane footprint. \\ Otherwise like draft\_connected.\end{tabular} &
  -''- &
  -''- \\\hline\\
S6-MS &
  light\_roll\_v2.99\_10yrs &
  \begin{tabular}[c]{@{}l@{}}Draft\_connected footprint but \\ full strength near-sun twilight \\ microsurvey. Rolling flipped, \\ roll strength 0.5, higher \\ suppress revisits weight, more \\ g in SCP/GP.\end{tabular} &
  -''- &
  -''- \\\hline
S7-MS &
  retro\_baseline\_v2.0\_10yrs &
  \begin{tabular}[c]{@{}l@{}}v1.x baseline with rubin\_sim \\ (without new code capabilities)\end{tabular} &
  baseline &
  \begin{tabular}[c]{@{}l@{}}The v2 baseline simulations. V2.0 and V2.1 feature\\ a survey footprint with expanded dust-free area and \\ WFD-level visits in the Galactic Bulge and Magellanic \\ Clouds. Coverage of the Northern Ecliptic Spur, South \\ Celestial Pole, and remainder of the Galactic Plane is \\ maintained, at lower levels. Filter balance is modified \\ in different areas of the sky. A 2-region rolling cadence \\ is used in the dust-free WFD sky, to improve cadence. \\ The v2.1 simulation adds coverage in the Virgo cluster \\ and acquisition of good seeing images in r \& i bands. \\ The retro simulations are intended to provide a bridge \\ from v1.X to v2.X, introducing modifications to the \\ footprint and scheduler code separately and in stages.\end{tabular} \\\hline\\
S8-MS &
  roll\_early\_v2.99\_10yrs &
  \begin{tabular}[c]{@{}l@{}}Starting rolling early and \\ increase suppress revisit \\ weight. DDF at 7.5\%, \\ twilight NEO microsurvey, \\ priority GP footprint.\end{tabular} &
  draft v3 &
  \begin{tabular}[c]{@{}l@{}}Draft versions of simulations incorporating the \\ SCOC Phase 2 recommendations for survey strategy.\end{tabular} \\ \hline\\\hline\\
S1-DDF &
  ddf\_accourd\_sf0.30\_lsf0.4\_lsr0.5\_v2.1\_10yrs &
  \begin{tabular}[c]{@{}l@{}}DDFs with an accordian \\ cadence: \\ 0.30 season fraction, \\ 0.4 low season fraction, \\ 0.5 low season rate\end{tabular} &
  ddf accordian &
  \begin{tabular}[c]{@{}l@{}}This family adds an "accordian" low/high rate over \\ each season approach to DDF cadence. The edges \\ of each season are observed at a "low" rate while \\ the center of the season is observed at a faster \\ "high" rate. Each simulation is identified by a total\\  season length fraction (as in `ddf season`) with a \\ similar low season fraction (`lsf`) within that and \\ a low season rate (lsr). Thus \\ `ddf\_accourd\_sf0.30\_lsf0.4\_lsr0.1\_v2.1\_10yrs` \\ would have a fairly short overall season (40\% of \\ the available season), a long low season (80\% - so\\  20\% high season), and a small low season rate (0.1), \\ leaving most DDF visits concentrated in the center \\ of the season. Conversely, \\ `ddf\_accourd\_sf0.10\_lsf0.1\_lsr0.5\_v2.1\_10yrs` \\ would have a long overall season, a short low season, \\ and a high rate during the low season, resulting in a \\ very mild accordian effect. This family uses \\ pre-scheduled DDF visits.\end{tabular} \\\hline
S2-DDF &
  ddf\_bright\_slf0.35\_v2.1\_10yrs &
  \begin{tabular}[c]{@{}l@{}}DDF quad-style sequences \\ running through bright time, \\ season length fraction 0.35\end{tabular} &
  ddf bright &
  \begin{tabular}[c]{@{}l@{}}Similar to the ddf quad family, the sequences for \\ individual DDFs are cut shorter by about 1/4 \\ compared to baseline. In this family, the m5 limits\\ for scheduling visits are relaxed, resulting in visits \\ running more consistently throughout the lunar \\ cycle even into bright time. This family uses \\ pre-scheduled DDF visits.\end{tabular} \\\hline\\
S3-DDF &
  ddf\_double\_slf0.35\_v2.1\_10yrs &
  \begin{tabular}[c]{@{}l@{}}Shorter (1/2) DDF sequences, \\ season length fraction 0.35\end{tabular} &
  ddf double &
  \begin{tabular}[c]{@{}l@{}}This family cuts the individual DDF sequences \\ in half, resulting in brighter single night coadded \\ depths, but a shorter interval of nights between \\ visits. This family uses pre-scheduled DDF visits.\end{tabular} \\\hline\\
S4-DDF &
  ddf\_old\_rot\_slf0.35\_v2.1\_10yrs &
  \begin{tabular}[c]{@{}l@{}}DDF with old rotation dithers \\ (constant rotTelPos), season \\ length fraction 0.35\end{tabular} &
  ddf old rot &
  \begin{tabular}[c]{@{}l@{}}This family adds a validation or comparison \\ option for the remainder of the v2.1 DDF \\ simulations. In this family, the rotation angle \\ is fixed so that rotTelPos is constant; in most \\ v2.1 simulations, rotSkyPos is held fixed. \\ Otherwise this family is similar to \\ `ddf season length`, and uses pre-scheduled \\ DDF visits.\end{tabular} \\\hline\\
S5-DDF &
  ddf\_quad\_slf0.35\_v2.1\_10yrs &
  \begin{tabular}[c]{@{}l@{}}Shorter (1/4) DDF sequences, \\ season length fraction 0.35\end{tabular} &
  ddf quad &
  \begin{tabular}[c]{@{}l@{}}This family cuts the individual DDF \\ sequences by four, resulting in brighter single \\ night coadded depths, but an even shorter \\ interval of nights between visits.This family \\ uses pre-scheduled DDF visits.\end{tabular} \\\hline
S6-DDF &
  ddf\_quad\_subfilter\_slf0.35\_v2.1\_10yrs &
  \begin{tabular}[c]{@{}l@{}}Shorter DDF sequences with \\ subsets of the filters, season \\ length fraction 0.35\end{tabular} &
  ddf quad subfilter &
  \begin{tabular}[c]{@{}l@{}}Similar to the ddf quad family, the sequences \\ for individual DDFs are cut shorter by about \\ 1/4 compared to baseline. However, here only \\ some filters are used in each DDF sequence, \\ alternating between active filters on different \\ nights.This family uses pre-scheduled DDF \\ visits.\end{tabular} \\\hline\\
S7-DDF &
  ddf\_season\_length\_slf0.20\_v2.1\_10yrs &
  \begin{tabular}[c]{@{}l@{}}DDF with season length \\ fraction 0.20\end{tabular} &
  ddf season length &
  \begin{tabular}[c]{@{}l@{}}This family investigates the effect of varying \\ the season length within the DDF fields. \\ `season length fraction` reflects how much of \\ the available season is used -- 0.1 indicates \\ that 10\% of the available season is removed \\ at each end (for a total of 80\% of the available \\ season used for observations), while 0.3 would \\ indicate that 40\% of the total possible season is \\ used. This family uses pre-scheduled DDF visits.\end{tabular} \\\hline

S8-DDF &
  ddf\_season\_length\_slf0.35\_v2.1\_10yrs &
  \begin{tabular}[c]{@{}l@{}}DDF with season length \\ fraction 0.35\end{tabular} &
  ddf season length &
  -''- \\ \hline

\end{longtable}
\end{landscape}
\clearpage
\twocolumn

\end{document}